\documentclass[showpacs,twocolumn,superscriptaddress]{revtex4}
\usepackage{doi}
\usepackage{hyperref}
\hypersetup{
  colorlinks=true,        
  linkcolor=blue,         
  citecolor=cyan,         
}
\usepackage[utf8]{inputenc}
\usepackage{graphicx}
\usepackage{dcolumn}
\usepackage{bm}
\usepackage{color}
\usepackage{enumitem}
\usepackage{amsmath}
\usepackage{amssymb}
\usepackage{orcidlink}
\usepackage{soul}

\begin{document}

\title{Effect of gravitational lensing around black hole in dark matter halo in the presence of plasma}
    \author{Zhiyu Dou}
    \affiliation{School of Physics, Harbin Institute of Technology, Harbin 150001, People’s Republic of China}

    \author{Akbar Davlataliev}
    \email{akbar@astrin.uz}
    \affiliation{School of Physics, Harbin Institute of Technology, Harbin 150001, People’s Republic of China}
    \affiliation{University of Tashkent for Applied Sciences, Str. Gavhar 1, Tashkent 100149, Uzbekistan}
    \affiliation{New Uzbekistan University, Movarounnahr str. 1, Tashkent 100000, Uzbekistan}
    
    \author{Mirzabek Alloqulov,\orcidlink{0000-0001-5337-7117}}
    \email{malloqulov@gmail.com}
    \affiliation{School of Physics, Harbin Institute of Technology, Harbin 150001, People’s Republic of China}
    \affiliation{Tashkent State Technical University, Tashkent 100095, Uzbekistan}
    \affiliation{New Uzbekistan University, Movarounnahr str. 1, Tashkent 100000, Uzbekistan}

	\author{Ahmadjon~Abdujabbarov,\orcidlink{0000-0002-6686-3787}}
	\email{ahmadjon@astrin.uz}

    \affiliation{School of Physics, Harbin Institute of Technology, Harbin 150001, People’s Republic of China}
    \affiliation{Institute of Theoretical Physics, National University of Uzbekistan, Tashkent 100174, Uzbekistan}

	\author{Bobomurat Ahmedov,\orcidlink{0000-0002-1232-610X}}
	\email{ahmedov@astrin.uz}
    \affiliation{School of Physics, Harbin Institute of Technology, Harbin 150001, People’s Republic of China}
	\affiliation{Institute of Theoretical Physics, National University of Uzbekistan, Tashkent 100174, Uzbekistan}
	\affiliation{Institute for Advanced Studies, New Uzbekistan University, Movarounnahr str. 1, Tashkent 100000, Uzbekistan}

    \author{Chengxun Yuan,\orcidlink{0000-0002-2308-6703}}
    \email{yuancx@hit.edu.cn}

    \affiliation{School of Physics, Harbin Institute of Technology, Harbin 150001, People’s Republic of China}

    \author{Chen Zhou,\orcidlink{0000-0002-1361-3717}}
\email{chenzhou@hit.edu.cn}

\affiliation{School of Physics, Harbin Institute of Technology, Harbin 150001, People’s Republic of China}


\date{\today}
\begin{abstract}
This article is devoted to the investigation of the observational properties of the Schwarzschild black hole (BH) surrounded by a dark matter (DM) halo. Our study commences with a brief review of spacetime, including the horizon structure and curvature invariants, which are the Ricci scalar, the square of the Ricci tensor, and the Kretschmann scalar. Subsequently, we explore the massive and massless particle dynamics around the Schwarzschild BH surrounded by a dark matter halo, including the innermost stable circular orbit (ISCO) and photon sphere radii. It was found that the radius of the ISCO increases under the influence of the spacetime parameters. Additionally, we investigate the weak gravitational lensing with the assumption that the BH is surrounded by a uniform and non-uniform plasma. Finally, we examine the impact of a plasma on the BH shadow and employ  Event Horizon Telescope (EHT) observational data to constrain the BH's parameters.
\end{abstract}

\maketitle

\section{Introduction}

Although General Relativity (GR) remains the cornerstone theory of gravitation and has passed numerous experimental tests in the weak-field regime—from solar system observations to binary pulsar timings—its suitability in extreme gravitational environments, such as those near black holes and in cosmology, continues to be scrutinized \citep{will_2014, berti_2015}. The quest for a unified theory of quantum gravity, the unexplained nature of dark energy and dark matter, and certain cosmological tensions motivate the exploration of alternative and modified theories of gravity \citep{clifton_2012, nojiri_2017}. Such theories often predict deviations from GR in the strong-field regime, which may be probed through astrophysical observations of compact objects, particularly black holes \citep{bambi_2017, cardoso_2019}.

In particular, black holes are not isolated entities but are embedded within complex astrophysical environments, often dominated by dark matter halos \citep{gondolo_1999, bertone_2005}. The presence of dark matter can significantly alter the spacetime geometry around black holes, influencing their observational signatures, such as quasinormal modes, shadows, and accretion dynamics \citep{konoplya_2019, xavier_2023, macedo_2024}. Recent advances in gravitational-wave astronomy by LIGO/Virgo \citep{ligo_2016, ligo_2017} and direct imaging by the Event Horizon Telescope \citep{eht_m87_2019, eht_sgr_2022} have opened new windows for examining these strong-field modifications. Understanding the interplay between dark matter distributions and black hole spacetimes is thus essential for interpreting current and future observational data \citep{kavanagh_2020, cardoso_2022}.

In this work, we investigate the effect of  gravitational lensing around a Schwarzschild black hole immersed in a Dehnen-type dark matter halo. Gravitational lensing, the phenomenon in which light rays are deflected from their original paths because of the curvature of spacetime, provides one of the most direct observational probes of gravitational fields across a wide range of scales. Its manifestations can be broadly classified into two regimes: strong-field lensing, which occurs in the immediate vicinity of compact objects such as black holes and neutron stars, and weak-field lensing, observed at larger impact parameters within galaxies and galaxy clusters. Each regime offers unique and complementary constraints on the underlying geometry and potential modifications to general relativity (GR)~\cite{16,Bozza:2002b,Bozza:2005a,Bozza2008lens,Tsp:2015a,Bin:2010a,Li2020}.

In astrophysical environments, black holes are enveloped by magnetized plasmas whose frequency-dependent refractive index modifies electromagnetic wave trajectories beyond pure geodesic motion. This plasma-lensing interplay is particularly crucial in modified gravity theories, where deviations from the Kerr metric couple with plasma dispersion to produce distinctive, frequency-dependent lensing signatures. Disentangling these effects requires a precise theoretical framework to interpret data from instruments such as the Event Horizon Telescope (EHT) and future observatories such as the ngVLA and SKA, especially since plasma effects can mimic or mask metric modifications~\cite{BisnovatyiKogan2010,Rogers2015,Perlick2017,Atamurotov2021}. The influence of plasma on weak gravitational lensing has been extensively analyzed within a variety of gravitational frameworks~\cite{Alloqulov20251,Meliyeva:2025pxg,Rahmatov:2025ari,Saydullayev:2025oop,Al-Badawi2024,Alloqulov2023,Jiang2024,Alloqulov2024,Al-Badawi20242,Alloqulov20242,Atamurotov2022,Sharipov2026461,Alloqulov20252,Khasanov_2025}.

We investigate gravitational lensing around a black hole embedded in a dark matter halo in a plasma environment, considering various configurations. The structure of the paper is organized as follows. In Sec.~\ref{sec:2}, we introduce the Schwarzschild black hole surrounded by a dark matter halo. Section~\ref{sec.3} is devoted to the dynamics of massive particles, while Sec.~\ref{sec.4} discusses photon motion in a plasma environment. In Sec.~\ref{sec.5}, we analyze weak gravitational lensing in plasma, and Sec.~\ref{sec.6} presents the magnification of gravitationally lensed images. The black hole shadow in plasma is examined in Sec.~\ref{sec.7}. Section~\ref{sec.8} focuses on parameter estimation, and finally, in Sec.~\ref{sec.9}, we summarize our main conclusions.

\section{Schwarzschild Black Hole in a Dark Matter Halo\label{sec:2}}

In realistic astrophysical environments, black holes are not isolated objects, but are expected to be embedded in galactic structures filled with dark matter. In order to model the influence of such environments, it is necessary to consider black hole spacetimes surrounded by a dark matter halo. In this work, we adopt the Dehnen-$(\alpha,\beta,\gamma)$ type density profile, which provides a flexible and widely used description of galactic dark matter distributions \cite{Dehnen:1993uh,2010gfe..book.....M,Al-Badawi_JCAP,Alloqulov2025EPJC}. This profile allows one to model different halo configurations through the appropriate choices of the parameters $(\alpha,\beta,\gamma)$.

Following the approach developed in \cite{Toshmatov:2025rln}, the spacetime geometry of a Schwarzschild black hole immersed in a spherically symmetric dark matter halo can be described, in Boyer–Lindquist coordinates, by the static and spherically symmetric line element

\begin{align}
    ds^{2} = -f(r)\,dt^{2} + \frac{dr^{2}}{f(r)} + r^{2} d\theta^{2} + r^{2}\sin^{2}\theta\, d\phi^{2}.
\end{align}

The presence of the surrounding dark matter distribution modifies the metric function $f(r)$ with respect to the standard Schwarzschild case. The explicit form of this modification depends on the density profile and structural parameters of the halo, such as the central density and the characteristic radius. As a result, the combined system provides a more realistic description of the gravitational field in galactic environments, allowing one to study the interplay between the black hole and its dark matter surroundings.

For the particular case of the Dehnen profile with $\gamma = 0$, the metric function takes the form \cite{Toshmatov:2025rln}

\begin{align}
    f(r) = \exp \left[-\frac{4 \pi  \rho_s r_s^{3} (2 r + r_s)}{3  (r + r_s)^{2}}\right] - \frac{2 M}{r},
\end{align}

where $\rho_s$ and $r_s$ denote the characteristic density and scale radius of the dark matter halo, respectively, $M$ is the black hole mass.

It is important to note that in the absence of dark matter, namely in the limit $\rho_s \to 0$ and $r_s \to 0$, the above expression reduces to

\begin{align}
    f(r) = 1 - \frac{2M}{r},
\end{align}
which corresponds exactly to the standard Schwarzschild solution. Therefore, the metric given above can be regarded as a natural generalization of the Schwarzschild spacetime that incorporates the gravitational influence of a surrounding dark matter halo. For simplicity, in the following analysis we set the black hole mass to (M = 1).

\begin{figure*}
\centering
\includegraphics[width=0.45\linewidth]{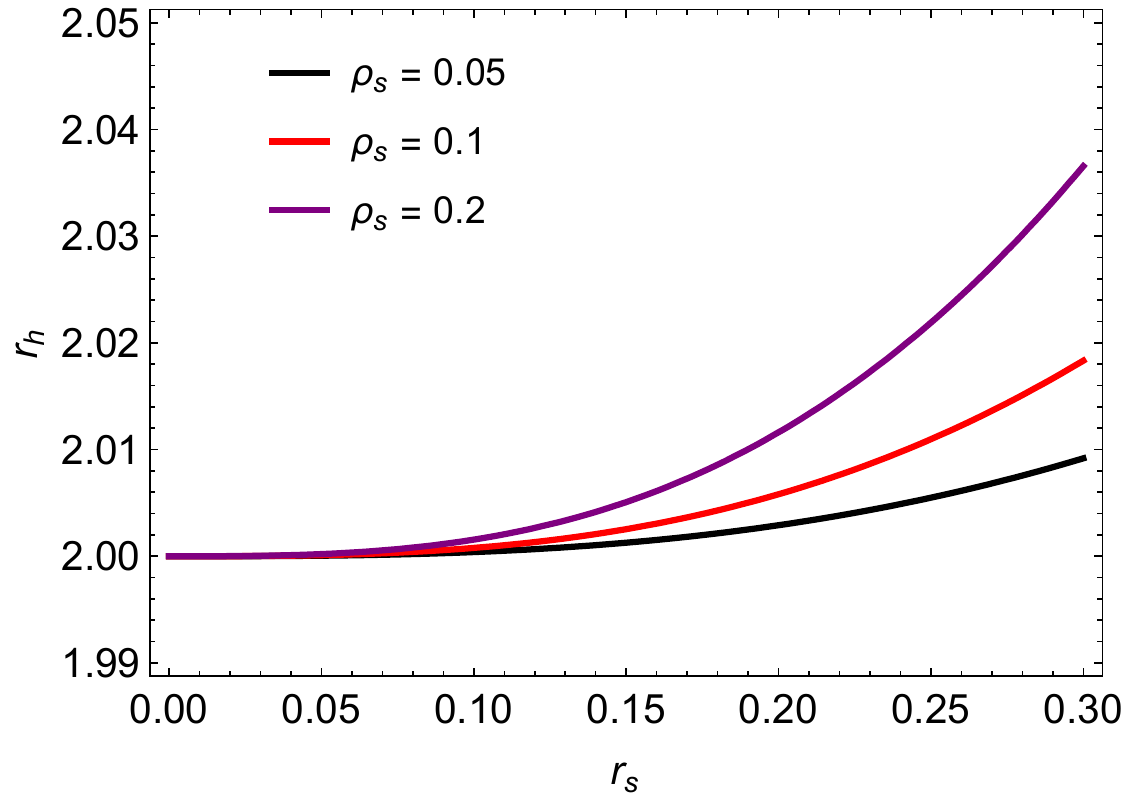}
\includegraphics[width=0.45\linewidth]{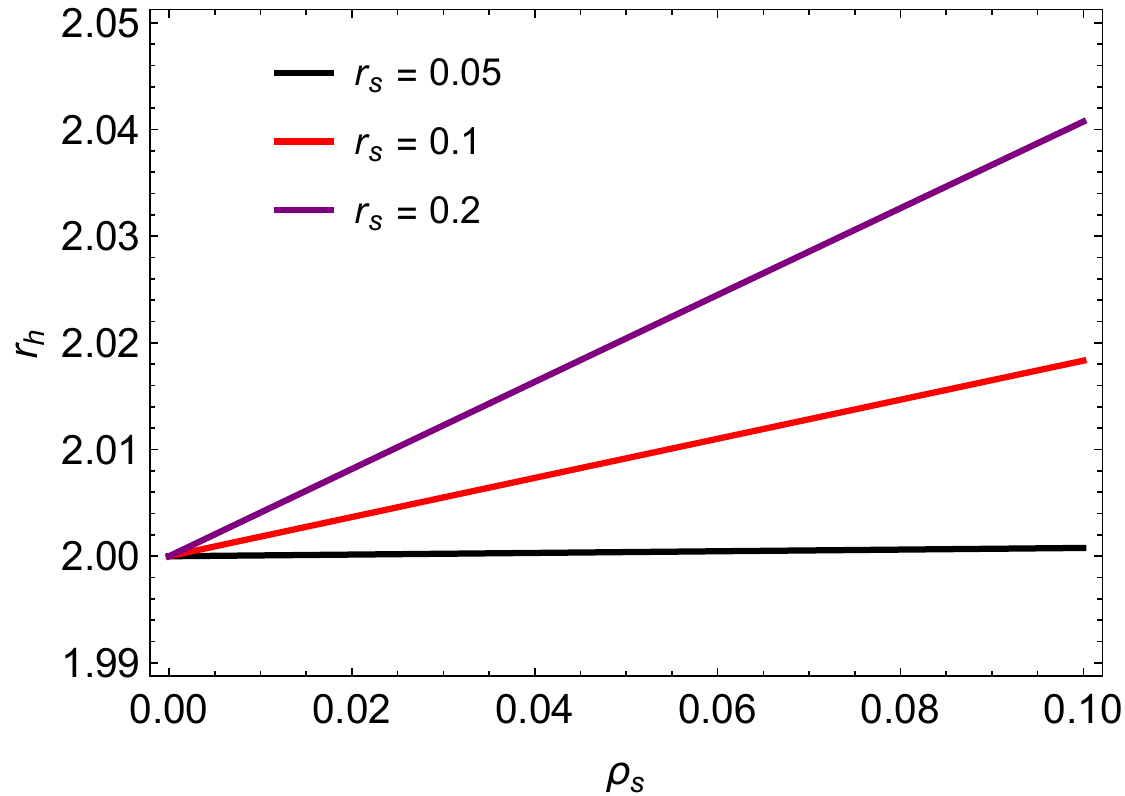}
\caption{Horizon radius as a function of the halo parameters $r_s$ and $\rho_s$.}
\label{fighorizon}
\end{figure*}

Figure~\ref{fighorizon} illustrates the behavior of the event horizon radius for a Schwarzschild black hole surrounded by a dark matter halo. It can be clearly seen that increasing the halo parameters leads to an enlargement of the horizon radius. In particular, higher values of $r_s$ and $\rho_s$ result in a larger black hole horizon, indicating that the presence of the dark matter halo modifies the spacetime geometry in such a way that the horizon expands.

\section{Dynamics of Massive Particles}\label{sec.3}

In order to study the motion of test particles around a Schwarzschild black hole surrounded by a dark matter halo, we begin with the standard Lagrangian formulation. The Lagrangian for a particle moving in a curved spacetime is given by

\begin{align}
    L = \frac{1}{2} g_{\mu\nu} u^{\mu} u^{\nu}, \qquad 
    u^{\mu} = \frac{dx^{\mu}}{d\tau},
\end{align}
where $u^{\mu}$ is the four-velocity of the particle and $\tau$ is the proper time.

Due to the spherical symmetry and stationarity of the spacetime, there exist two conserved quantities associated with the timelike and rotational Killing vectors. These correspond to the specific energy $\mathcal{E}$ and the specific angular momentum $\mathcal{L}$ of the particle. They are obtained from

\begin{align}
    \mathcal{E} = -\frac{\partial L}{\partial u^{t}} = f(r)\frac{dt}{d\tau},
\end{align}

\begin{align}
    \mathcal{L} = \frac{\partial L}{\partial u^{\phi}} = r^{2}\sin^{2}\theta\, \frac{d\phi}{d\tau}.
\end{align}

Without loss of generality, the motion can be restricted to the equatorial plane $\theta=\pi/2$. Using the normalization condition $g_{\mu\nu}u^{\mu}u^{\nu}=-1$ for massive particles, the radial equation of motion can be written as

\begin{align}
    \left(\frac{dr}{d\tau}\right)^{2} = \mathcal{E}^{2} - f(r)\left(1+\frac{\mathcal{L}^{2}}{r^{2}}\right).
\end{align}

This equation allows us to define the potential for massive particles as

\begin{align}
    V(r) = f(r)\left(1+\frac{\mathcal{L}^{2}}{r^{2}}\right).
\end{align}

For the metric function corresponding to a Schwarzschild black hole embedded in a dark matter halo, the potential takes the explicit form

\begin{align}
V(r)=\left(1+\frac{\mathcal{L}^{2}}{r^{2}}\right)
\left\{
\exp \left[-\frac{4 \pi \rho_s r_s^3 (2 r+r_s)}{3 (r+r_s)^2}\right]
-\frac{2}{r}
\right\}.
\end{align}

\begin{figure*}
    \centering
    \includegraphics[width=0.45\linewidth]{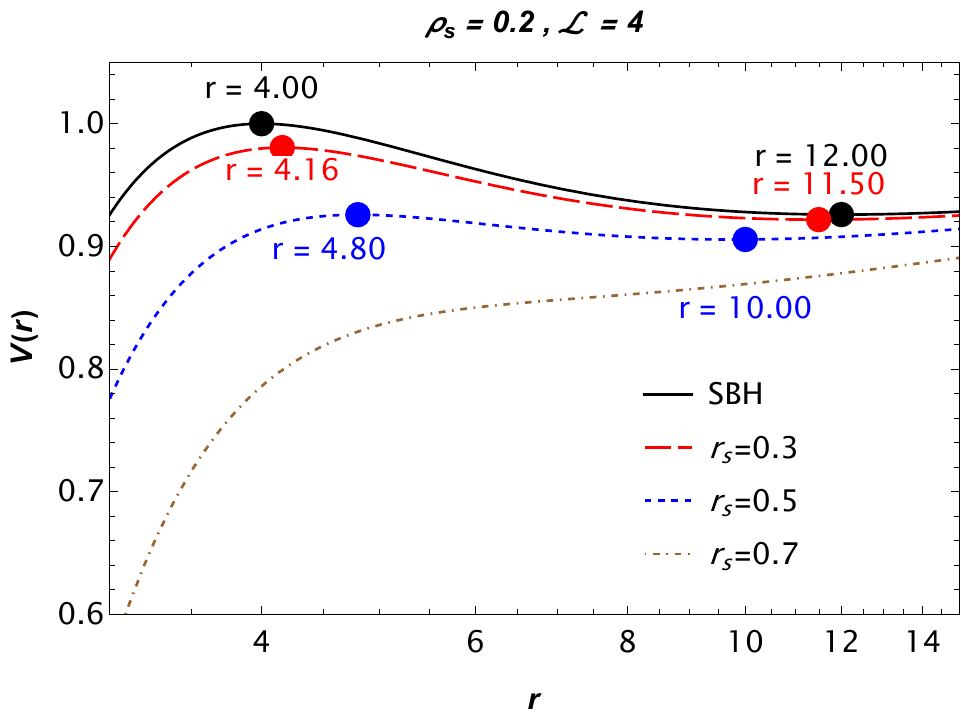}
    \includegraphics[width=0.45\linewidth]{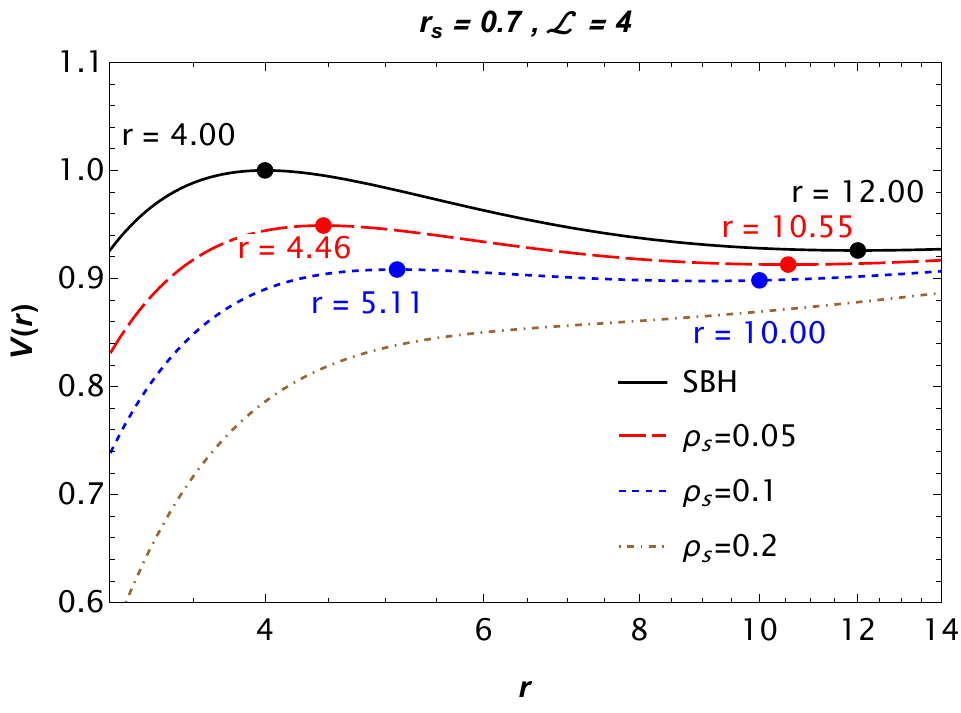}
    \caption{Radial dependencies of the potential $V(r)$ as a function of the halo parameters $r_s$ (left panel) and $\rho_s$ (right panel).}
    \label{figpotential}
\end{figure*}

The behavior of the effective potential as a function of the halo parameters $r_s$ and $\rho_s$ is illustrated in Fig.~\ref{figpotential}. These plots demonstrate how the presence of dark matter modifies the gravitational potential experienced by a massive particle; the locations of the maximum and minimum points (corresponding to unstable and stable circular orbits, respectively) are also presented. Furthermore, the effect of the halo parameters is easily seen by comparison with the Schwarzschild black hole (SBH) case, where $\rho_s = 0$. Increasing either $r_s$ or $\rho_s$ causes the maximum and minimum points of the effective potential to move toward each other.

By analyzing the potential, one can determine the allowed regions of motion as well as the energy and angular momentum of particles moving on circular orbits. For convenience, we introduce the following notation:

\begin{align}
\mathcal{E}^2 = \frac{9(r+r_s)^3 \left(A-\frac{2}{r}\right) \left(2 A-r\right)}{4 \pi \rho_s r^3 r_s^3 + 9 (r+r_s)^3A - 3 r (r+r_s)^3},
\end{align}
where
\begin{align}
    A = \exp \left[-\frac{4 \pi \rho_s r_s^3 (2 r+r_s)}{3 (r+r_s)^2}\right].
\end{align}

\begin{figure*}
    \centering
    \includegraphics[width=0.45\linewidth]{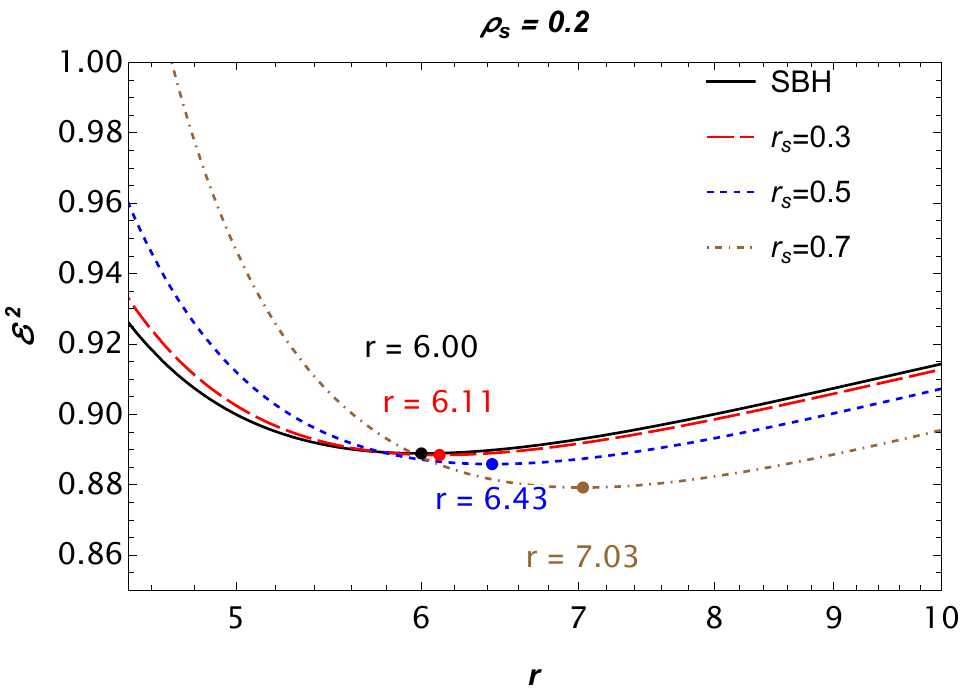}
    \includegraphics[width=0.45\linewidth]{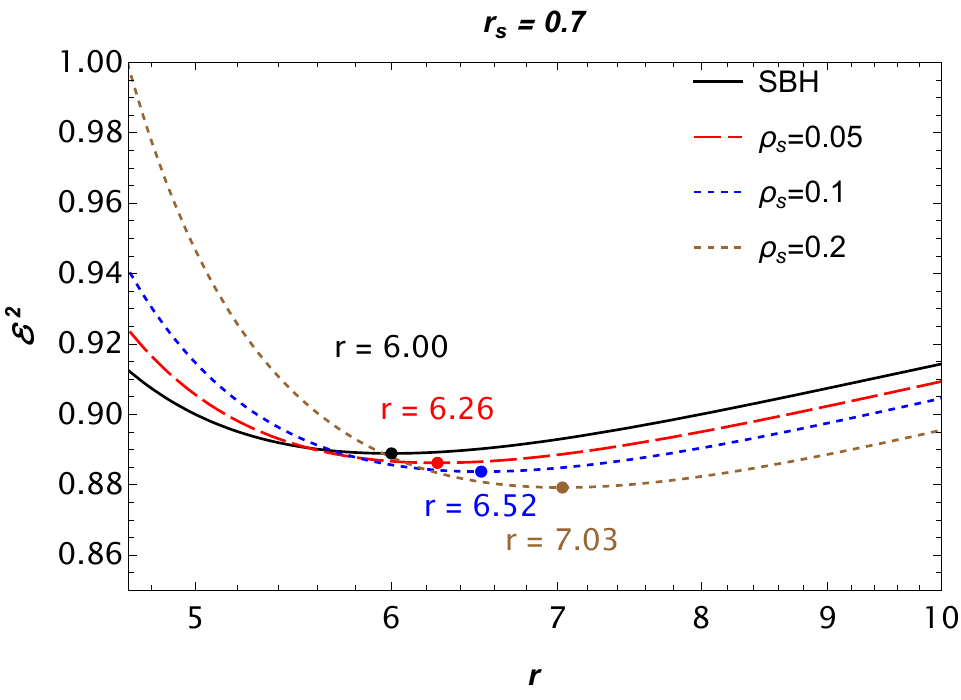}
    \caption{Radial dependencies of square of the specific energy $\mathcal{E}^2$ as a function of $r_s$ (left) and $\rho_s$ (right).}
    \label{figsqenergy}
\end{figure*}

Fig.~\ref{figsqenergy} shows the dependence of the specific energy on the halo parameters. The minima of the energy curves are indicated by filled circles, and the corresponding radial coordinates are explicitly specified. The minimum points of the specific energy shift to higher radii compared to the SBH case. 

Similarly, the square of the specific angular momentum of a particle on a circular orbit is given by

\begin{align}
\mathcal{L}^2 = \frac{r^2 \left(4 \pi \rho_s r^3 r_s^3 + 3 (r+r_s)^3 A\right)}{-4 \pi \rho_s r^3 r_s^3 - 9 (r+r_s)^3 A + 3 r (r+r_s)^3}.
\end{align}

\begin{figure*}
    \centering
    \includegraphics[width=0.45\linewidth]{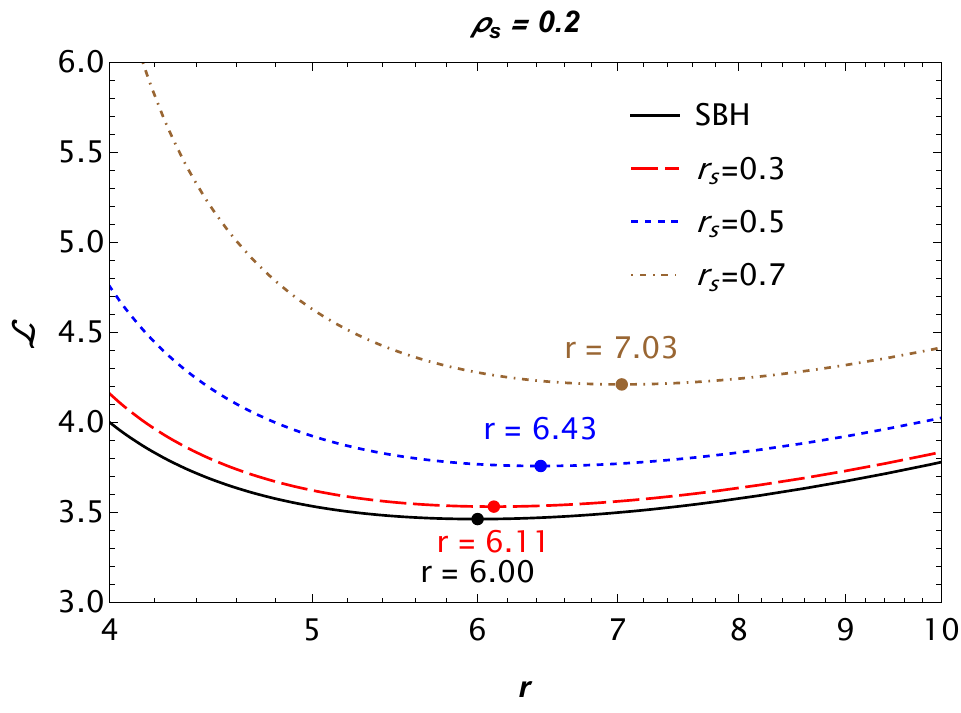}
    \includegraphics[width=0.45\linewidth]{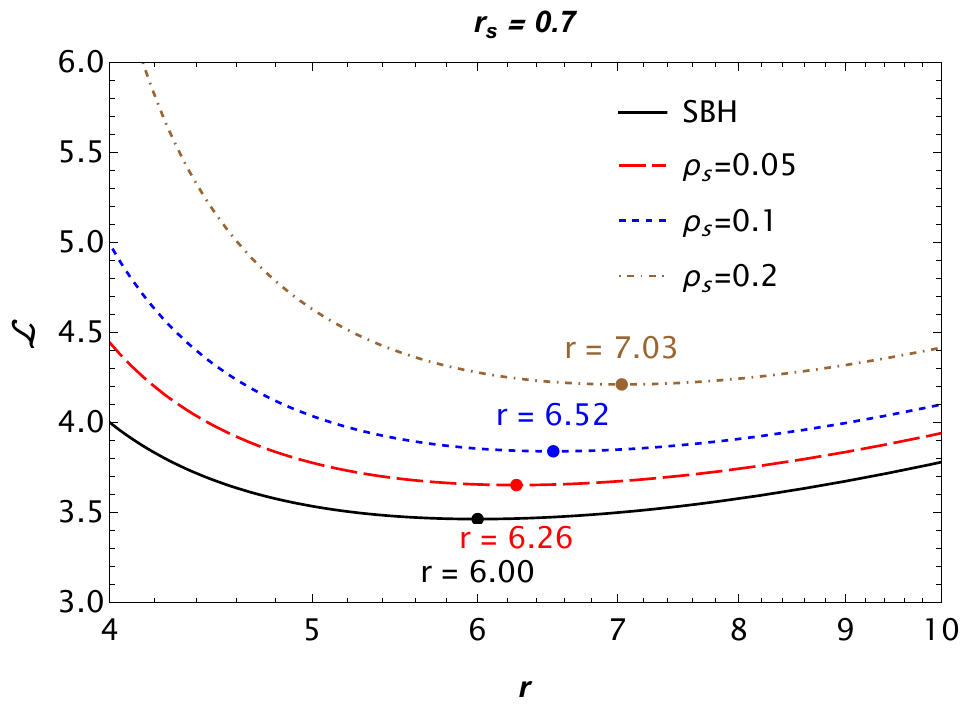}
    \caption{Radial dependencies of the specific angular momentum $\mathcal{L}$ as a function of $r_s$ (left) and $\rho_s$ (right).}
    \label{figsqangular}
\end{figure*}

Its behavior as a function of the halo parameters is presented in Fig.~\ref{figsqangular}. The minima of the profile are marked, together with their corresponding radial coordinates. Similarly to the trend observed for the specific energy, as the halo parameters $r_s$ and $\rho_s$ increase, the positions of the minima shift toward larger radial distances.

\subsection{Innermost Stable Circular Orbit (ISCO)}

The radius of the innermost stable circular orbit (ISCO) is one of the most important characteristics of a black hole spacetime, as it determines the inner edge of accretion disks and plays a key role in many astrophysical processes.

The ISCO radius is obtained from the standard conditions

\begin{align}
\frac{dV(r)}{dr}=0, \qquad 
\frac{d^{2}V(r)}{dr^{2}}=0.
\end{align}

\begin{figure*}
    \centering
    \includegraphics[width=0.45\linewidth]{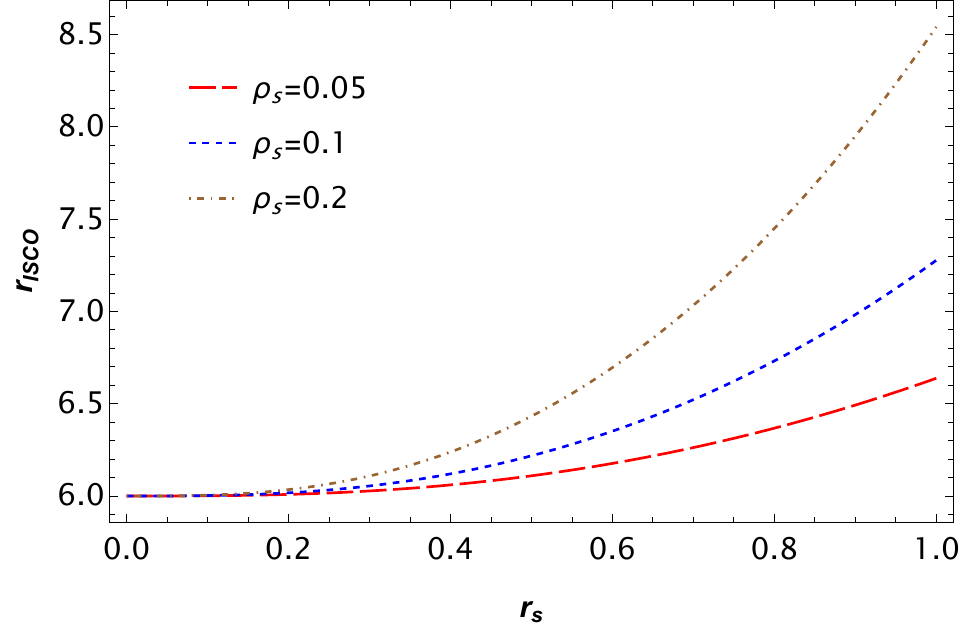}
    \includegraphics[width=0.45\linewidth]{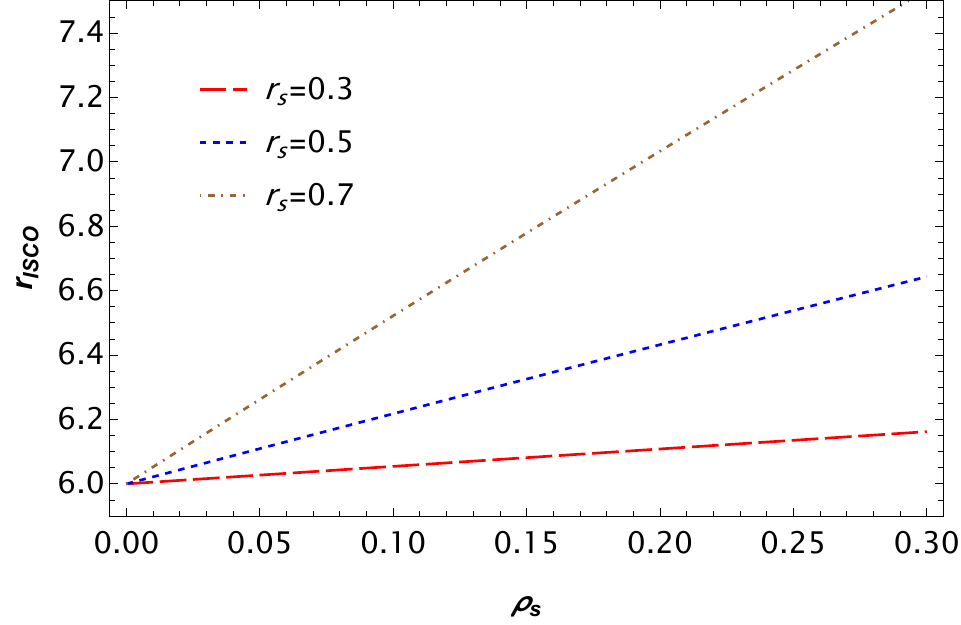}
    \caption{Dependence of the ISCO radius $R_{\text{ISCO}}$ on the halo parameters $r_s$ (left) and $\rho_s$ (right).} 
    \label{figISCO}
\end{figure*}

Solving these equations numerically for the metric under consideration yields the dependence of $r_{\rm ISCO}$ on the dark matter halo parameters. The results are presented in Fig.~\ref{figISCO}, which shows how the presence of dark matter can significantly shift the location of the ISCO compared to the standard Schwarzschild case.

\section{massless particle in plasma environment}\label{sec.4}

In a plasma environment, the Hamiltonian governing photon dynamics takes the following form \cite{Synge:1960ueh}:

\begin{equation}\label{hamilton_pl}
    H(x^{\alpha}, p_{\alpha}) = \frac{1}{2} \, \tilde{g}^{\alpha\beta} p_{\alpha} p_{\beta},
\end{equation}

\noindent where \( x^{\alpha} \) represent the spacetime coordinates and the effective metric \(\tilde{g}^{\alpha\beta}\) is expressed as

\begin{equation}
    \tilde{g}^{\alpha\beta} = g^{\alpha\beta} - (n^{2} - 1) u^{\alpha} u^{\beta}.
\end{equation}

\noindent Here, \( n \) stands for the refractive index of the plasma, while \( p_{\alpha} \) and \( u^{\beta} \) correspond to the photon's four-momentum and four-velocity, respectively. The refractive index is given by \cite{Mendonca:2019eke}

\begin{equation}\label{nindex}
    n^{2} = 1 - \frac{\omega_{p}^{2}(r)}{\omega^{2}(r)},
\end{equation}

\noindent with the electron plasma frequency defined as \( \omega_{p}^{2}(r) = 4\pi e^{2} N(r) / m_{e} \), where \( e \) and \( m_{e} \) are the electron charge and mass, and \( N(r) \) is the electron number density. The photon frequency \( \omega(r) \), as registered by a static observer, follows from the gravitational redshift relation:

\begin{equation}
    \omega(r) = \frac{\omega_{0}}{\sqrt{f(r)}}.
\end{equation}

\noindent The constant \( \omega_{0} \) signifies the photon frequency at infinity (\( f(\infty) = 1 \)), equivalent to the photon's energy at spatial infinity, \( \omega_{0} = \omega(\infty) = -p_{t} \) \cite{Perlick:2015vta}. Propagation within the plasma requires the photon frequency to surpass the plasma frequency, i.e., \( \omega_{p} / \omega < 1 \); otherwise, transmission ceases. When \( \omega_{0} \approx \omega_{p} \), the resulting deflection angle \( \alpha \) substantially exceeds the vacuum case (\( \omega_{p} = 0 \)), satisfying \( \alpha \gg 2R/b \), where \( b \) is the impact parameter.

Applying the canonical relation \( \dot{x}^{\alpha} = \partial H / \partial p_{\alpha} \) in conjunction with equations \eqref{hamilton_pl} and \eqref{nindex}, and restricting motion to the equatorial plane (\( \theta = \pi/2, \, p_{\theta} = 0 \)), the four-velocity components for photons become:

\begin{align}
    \dot{t} \equiv \frac{dt}{d\lambda} &= -\frac{p_{t}}{f(r)}, \label{eq:t_dot} \\
    \dot{r} \equiv \frac{dr}{d\lambda} &= p_{r} f(r), \label{eq:r_dot} \\
    \dot{\varphi} \equiv \frac{d\varphi}{d\lambda} &= \frac{p_{\varphi}}{r^{2}}. \label{eq:phi_dot}
\end{align}

\noindent Combining equations (\ref{eq:r_dot}) and (\ref{eq:phi_dot}) yields the phase trajectory of light:

\begin{equation}
    \frac{dr}{d\varphi} = \frac{\dot{r}}{\dot{\varphi}} = \frac{f(r) r^{2} p_{r}}{p_{\varphi}}.
\end{equation}

\noindent Imposing the null-condition \( H = 0 \) for photon trajectories allows this equation to be recast as

\begin{equation}
    \frac{dr}{d\varphi} = \pm r \sqrt{f(r)} \sqrt{ h^{2}(r) \frac{\omega_{0}^{2}}{p_{\varphi}^{2}} - 1 },
\end{equation}

\noindent where the function \( h^{2}(r) \) is defined as \cite{Perlick:2015vta}

\begin{equation}\label{funchr}
    h^{2}(r) \equiv r^{2} \left[ \frac{1}{f(r)} - \frac{\omega_{p}^{2}(r)}{\omega_{0}^{2}} \right].
\end{equation}

\noindent The radius \( r_{ph} \) of a circular photon orbit, which constitutes the photon sphere, is found by solving the critical condition \cite{Perlick:2021aok}

\begin{equation}\label{derhr}
    \left. \frac{d\bigl( h^{2}(r) \bigr)}{dr} \right|_{r=r_{ph}} = 0.
\end{equation}

\noindent Substituting equation \eqref{funchr} into \eqref{derhr} produces an algebraic equation determining \( r_{p} \) in the presence of a plasma:

\begin{align}\label{eqphsph}
    \frac{2 f(r)-rf'(r)}{2f(r)^2}\Bigg|_{r=r_{ph}}=\frac{\omega^2_p(r)}{\omega_0^2}+\frac{r\omega_p(r)\omega_p'(r))}{\omega_0^2}\Bigg|_{r=r_{ph}}
\end{align}

\noindent Here, the prime denotes differentiation with respect to \( r \). Analytical solutions for \( r_{ph} \) are generally unavailable for arbitrary \( \omega_{p}(r) \) profiles; consequently, specific simplified cases will be examined in the subsequent analysis.

\subsection{Homogeneous Plasma}

\begin{figure*}
    \centering
    \includegraphics[width=0.45\linewidth]{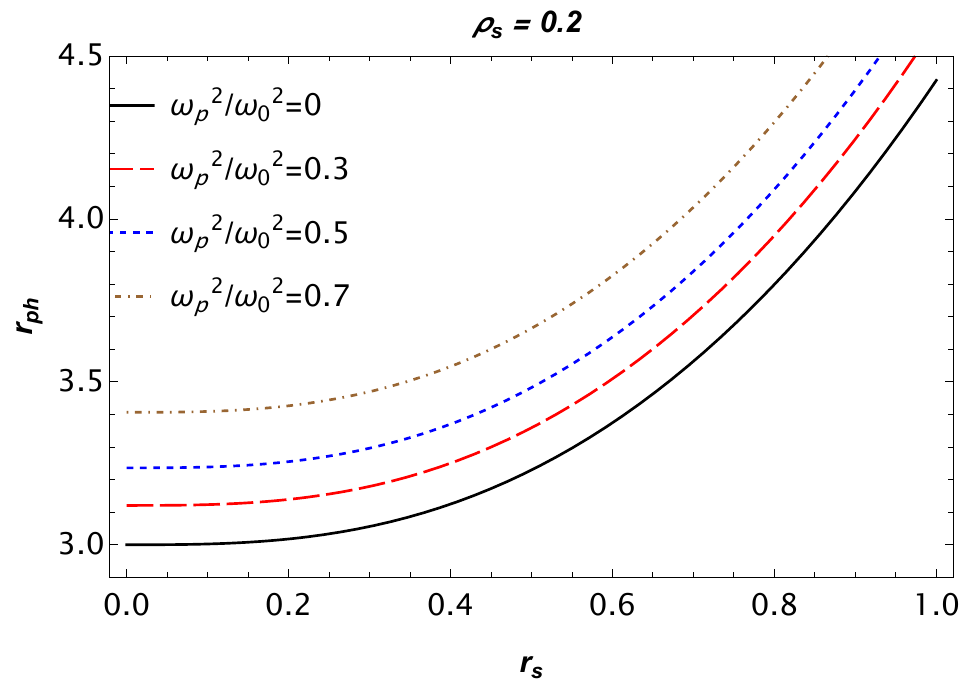}
    \includegraphics[width=0.45\linewidth]{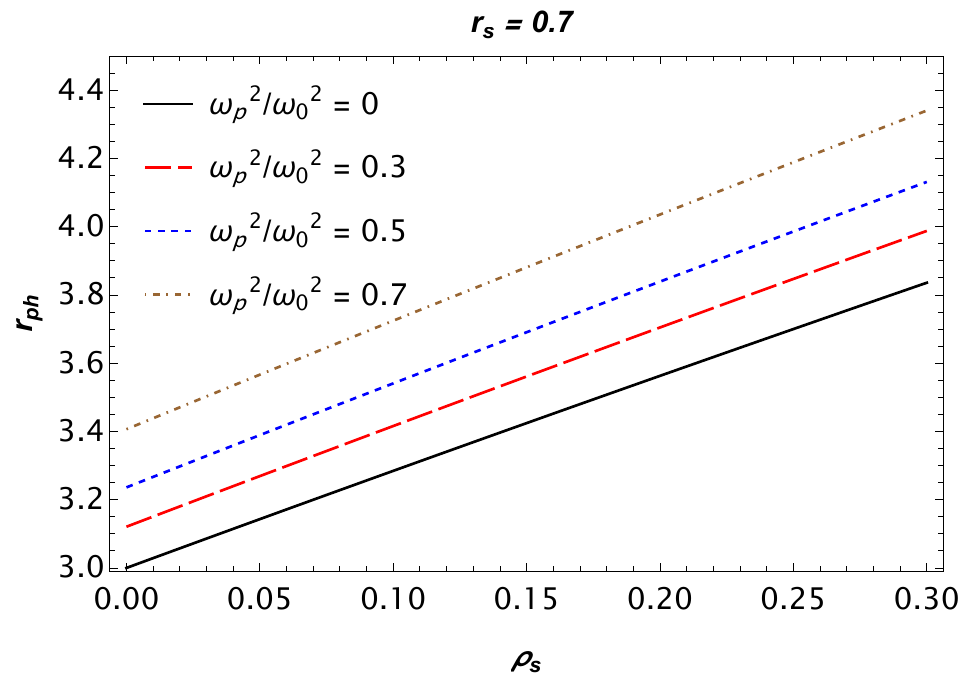}
   \caption{Photon sphere radius as a function of the dark matter parameters $r_{s}$ (left panel) and $\rho_{s}$ (right panel) for different values of the homogeneous plasma frequency.}
    \label{homophsph}
\end{figure*}

First, we consider the case of a homogeneous plasma medium, where the plasma frequency \(\omega_{p}^{2}(r)\) is constant throughout the region. Under this condition, Eq.~\eqref{eqphsph} can be solved numerically, and the results are illustrated in Fig.~\ref{homophsph}. From this figure, it can be observed that the photon sphere radius \(r_{ph}\) increases as the dark matter halo parameters $r_s$ and $\rho_s$ increases. Additionally, the presence of a plasma medium tends to increase the radius of the photon sphere compared to the vacuum case.

\subsection{Inhomogeneous Plasma}

\begin{figure*}
    \centering
    \includegraphics[width=0.45\linewidth]{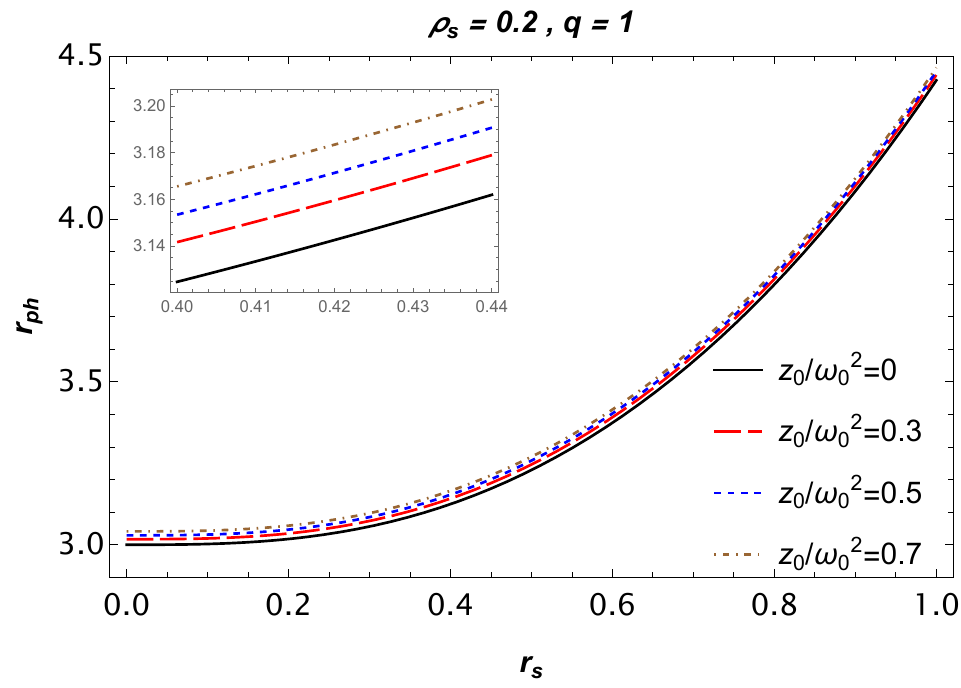}
    \includegraphics[width=0.45\linewidth]{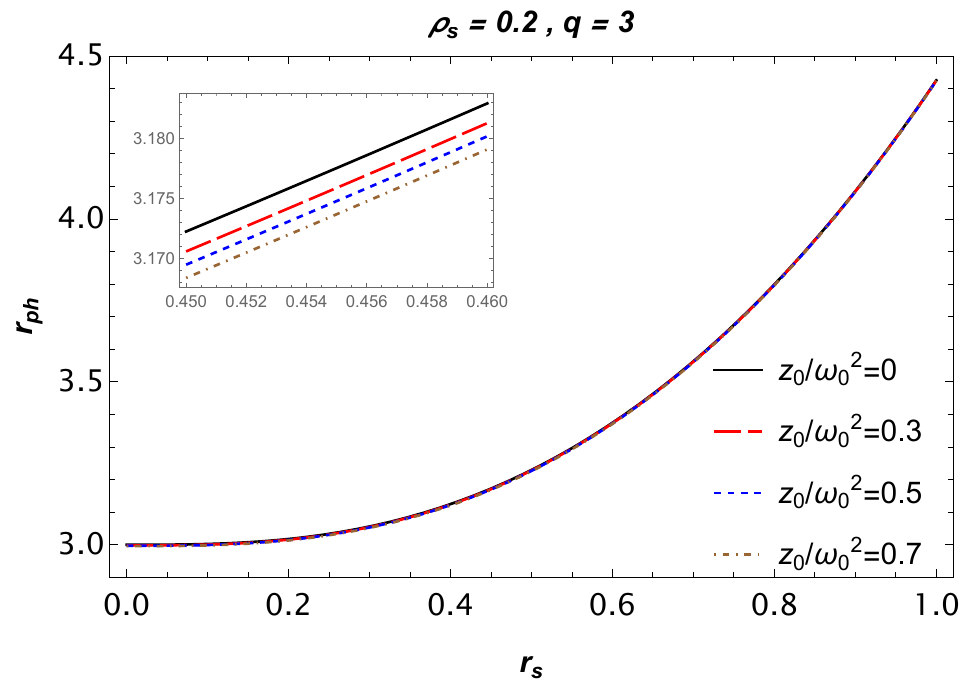}
    \includegraphics[width=0.45\linewidth]{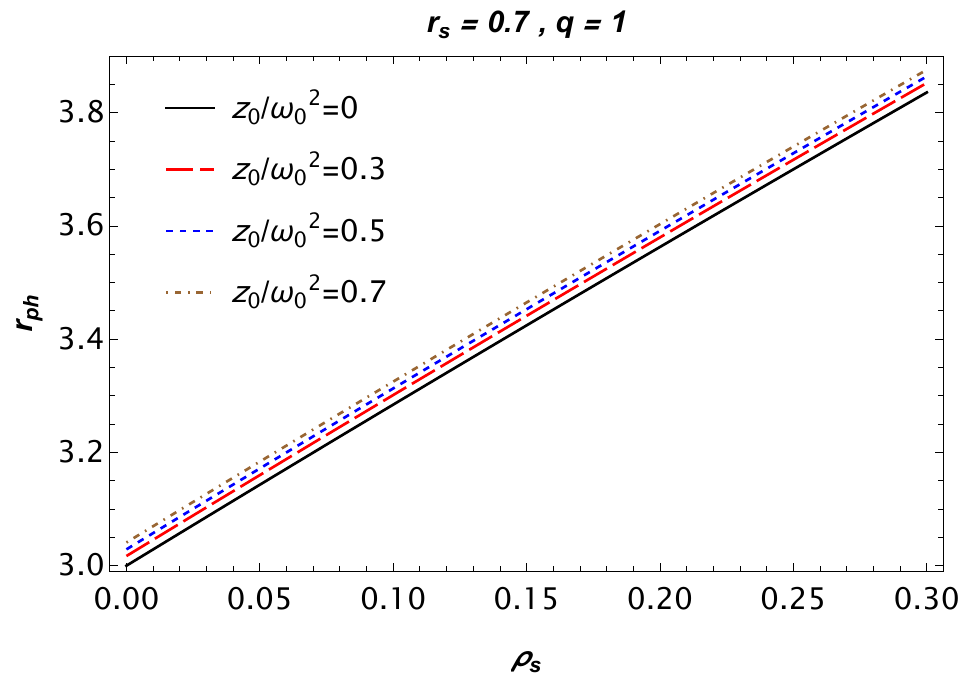}
    \includegraphics[width=0.45\linewidth]{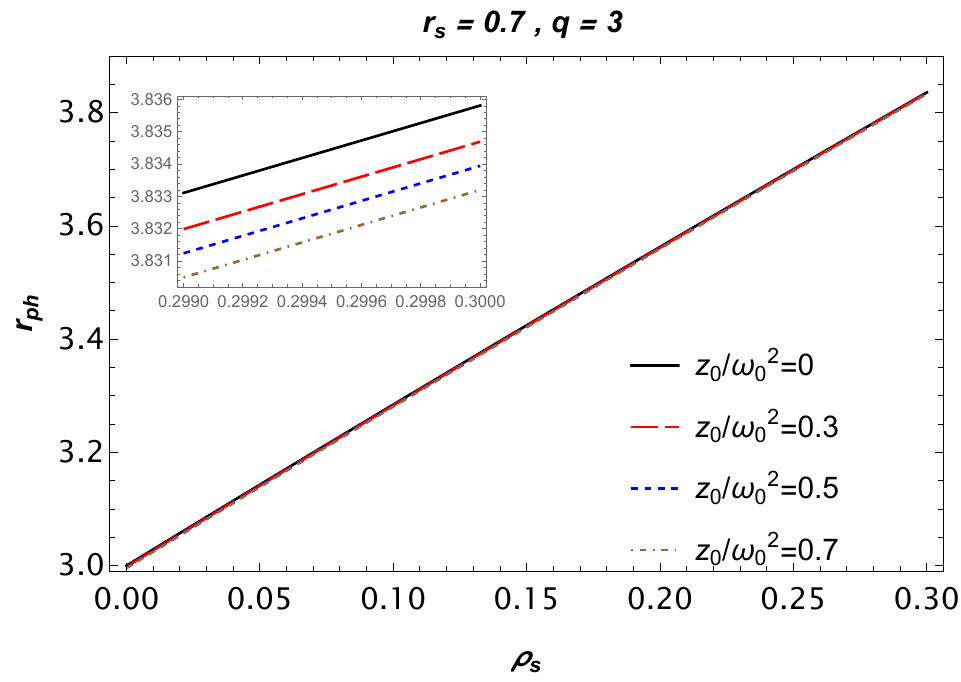}
    \caption{Photon sphere radius as a function of the dark matter parameters $r_{s}$ (left panel) and $\rho_{s}$ (right panel) for different values of the inhomogeneous plasma frequency.}
    \label{nonhomophsph}
\end{figure*}

This subsection investigates photon spheres in the context of a non-uniform plasma distribution. We assume the plasma frequency follows a simple power-law profile of the form \cite{Rogers:2015dla,Er:2017lue},

\begin{equation}
\omega_{p}^{2} (r) = \frac{z_{0}}{r^{q}},
\label{eq:plasma_power_law}
\end{equation}
where \(z_{0}\) and \(q > 0\) are free parameters. To explore the fundamental characteristics of this model, we focus on two specific cases:
\begin{itemize}
    \item \(q = 1\) with constant \(z_{0}\), which precisely replicates the negative-mass diverging lens model \cite{Er:2017lue}.
    \item \(q = 3\) with constant \(z_{0}\), a profile associated with the stellar surface based on the Goldreich-Julian (GJ) density.
\end{itemize}
By combining Equations~\eqref{eqphsph} and (\ref{eq:plasma_power_law}), the radius of the photon sphere for an inhomogeneous plasma medium is determined via a numerical scheme. The results of this analysis are presented in Fig.~\ref{nonhomophsph}. We observe that the halo parameters $r_s$ and $\rho_s$ have the same qualitative effect: increasing either of them leads to a larger photon sphere radius $r_{ph}$. Additionally, the parameter $q$ exhibits an inverse behavior; specifically, the case $q = 3$ produces the opposite trend compared to $q = 1$.

\section{Weak-Field Gravitational Lensing in a Plasma Environment}\label{sec.5}

This section is dedicated to investigating the effects of gravitational lensing for a Schwarzschild black hole embedded within a dark matter halo and surrounded by a plasma medium. The analysis employs the weak-field approximation, where the spacetime metric is expressed as a perturbation around the Minkowski background \cite{Bisnovatyi-Kogan:2010flt}:
\begin{align}
    g_{\alpha\beta} = \eta_{\alpha\beta} + h_{\alpha\beta},
\end{align}
where \(\eta_{\alpha\beta}\) is the flat Minkowski metric, \(h_{\alpha\beta}\) is a small perturbation, and the following conditions hold:
\begin{align}
    \eta_{\alpha\beta} &= \text{diag}(-1, 1, 1, 1),\nonumber \\
    h_{\alpha\beta} &\ll 1, \quad \text{and} \quad h_{\alpha\beta} \to 0 \quad \text{as} \quad x^\alpha \to \infty.
\end{align}
The inverse metric, to first order, is given by \(g^{\alpha\beta} = \eta^{\alpha\beta} - h^{\alpha\beta}\), with \(h^{\alpha\beta} = h_{\alpha\beta}\).

Within this framework, the deflection angle \(\hat{\alpha}_b\) for a light ray with impact parameter \(b\) can be derived as \cite{Bisnovatyi-Kogan:2010flt}:
\begin{widetext}
\begin{align}
    \hat{\alpha}_b = \frac{1}{2} \int_{-\infty}^{\infty} \frac{b}{r} \left( \frac{dh_{33}}{dr} + \frac{1}{1 - \omega_p^2 / \omega^2} \frac{dh_{00}}{dr} - \frac{K_e}{\omega^2 - \omega_p^2} \frac{dN}{dr} \right) dz.
    \label{eq:deflection_angle_full}
\end{align}
\end{widetext}
Here, \(N(x^i)\) denotes the number density of plasma particles around the black hole, and \(K_e = 4\pi e^2 / m_e\) is a constant (with \(m_e\) being the electron mass).

In Cartesian coordinates, with \(r^2 = b^2 + z^2\), the metric perturbations for this specific spacetime configuration are:
\begin{align}
    h_{00} &= 1 + \frac{2}{r} - A, \nonumber\\
    h_{ik} &= \left( 1 + \frac{2}{r} - A \right) n_i n_k\ ,\nonumber \\
    h_{33} &= \left( 1 + \frac{2}{r} - A \right) \cos^2 \chi,
\end{align}
where \(\cos^2 \chi = z^2 / (b^2 + z^2)\) and \(n_i\) are the components of the unit radial vector.

The total deflection angle in Equation~\eqref{eq:deflection_angle_full} naturally decomposes into three distinct components, each associated with a specific physical contribution:
\begin{align}\label{sumdefangle}
    \hat{\alpha}_{uni} &= \hat{\alpha}_{uni1} + \hat{\alpha}_{uni2} + \hat{\alpha}_{uni3}\ ,\nonumber \\
    \hat{\alpha}_{uni1} &= \frac{1}{2} \int_{-\infty}^{\infty} \frac{b}{r} \frac{dh_{33}}{dr} \, dz\ ,\nonumber \\
    \hat{\alpha}_{uni2} &= \frac{1}{2} \int_{-\infty}^{\infty} \frac{b}{r} \frac{1}{1 - \omega_p^2 / \omega^2} \frac{d h_{00}}{dr} \, dz\ ,\nonumber \\
    \hat{\alpha}_{uni3} &= \frac{1}{2} \int_{-\infty}^{\infty} \frac{b}{r} \left( -\frac{K_e}{\omega^2 - \omega_p^2} \frac{dN}{dr} \right) dz=0.
\end{align}
Here, \(\hat{\alpha}_1\) arises from the spatial curvature perturbation, \(\hat{\alpha}_2\) from the modified time-time component of the metric weighted by the plasma dispersion, and \(\hat{\alpha}_3\) from the direct refractive effect of the plasma density gradient.

\subsection{Uniform Plasma}

For the first scenario, we examine a homogeneous plasma medium characterized by a constant plasma frequency, \(\omega_{p}^{2} = \text{const}\). Under this condition, the refractive index is spatially uniform and does not depend explicitly on the coordinates. Consequently, the refractive contribution associated with the plasma density gradient can be neglected. This implies that the final term in eq.~\eqref{sumdefangle}, denoted as \(\hat{\alpha}_3\), is not considered. Performing the integration of eq.~\eqref{sumdefangle} under this simplification yields the following expression for the deflection angle:

\begin{align}\label{eqhomoangle}
    \hat\alpha_{uni}=\hat\alpha_{uni1}+\hat\alpha_{uni2}
\end{align}
\begin{figure*}
    \centering
    \includegraphics[width=0.45\linewidth]{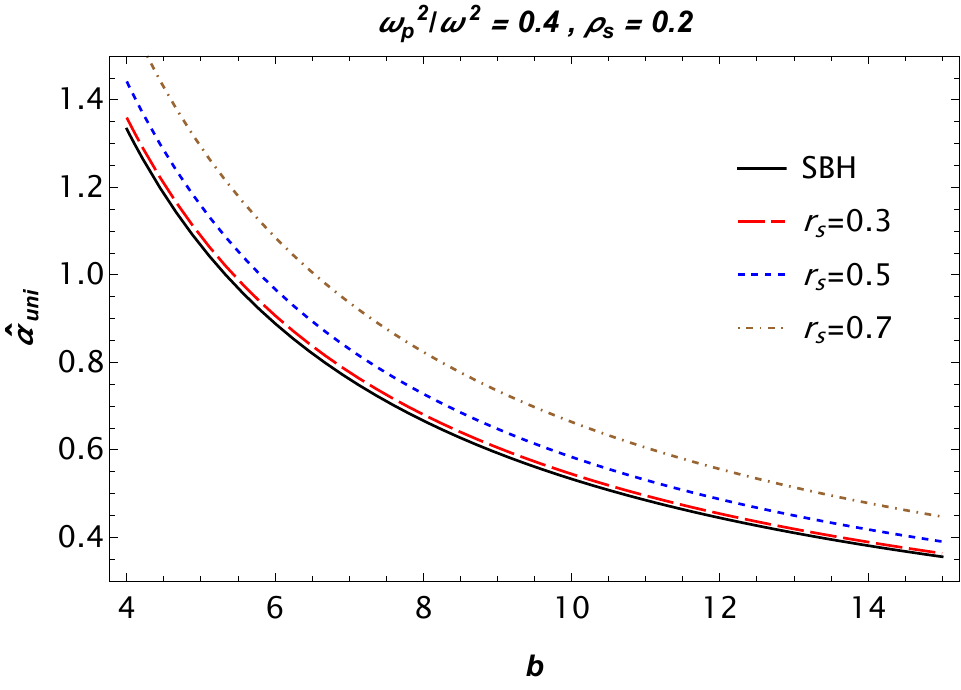}
    \includegraphics[width=0.45\linewidth]{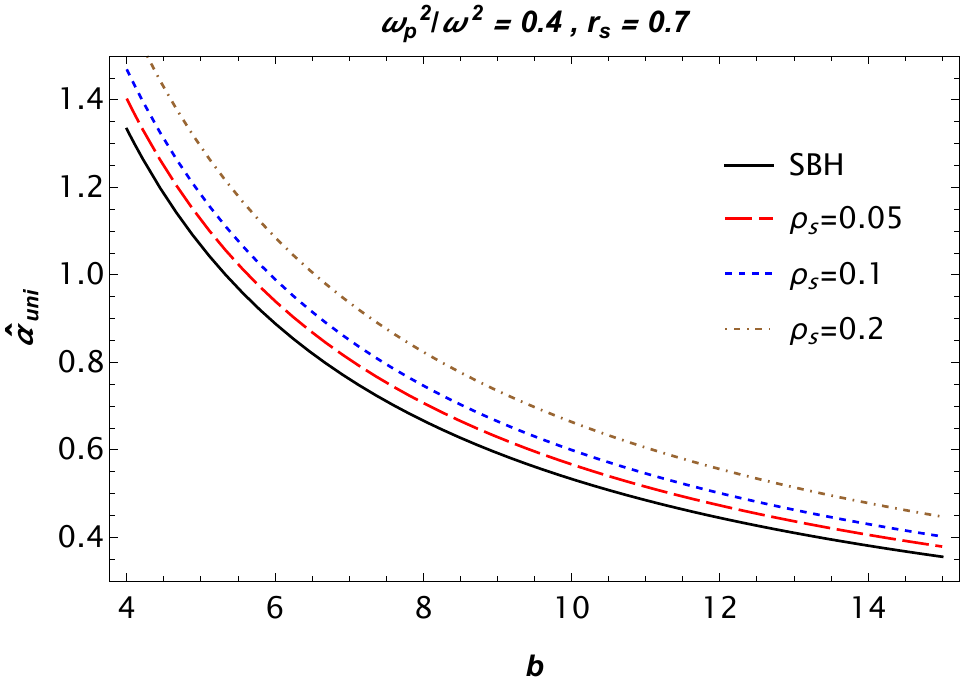}
  \includegraphics[width=0.45\linewidth]{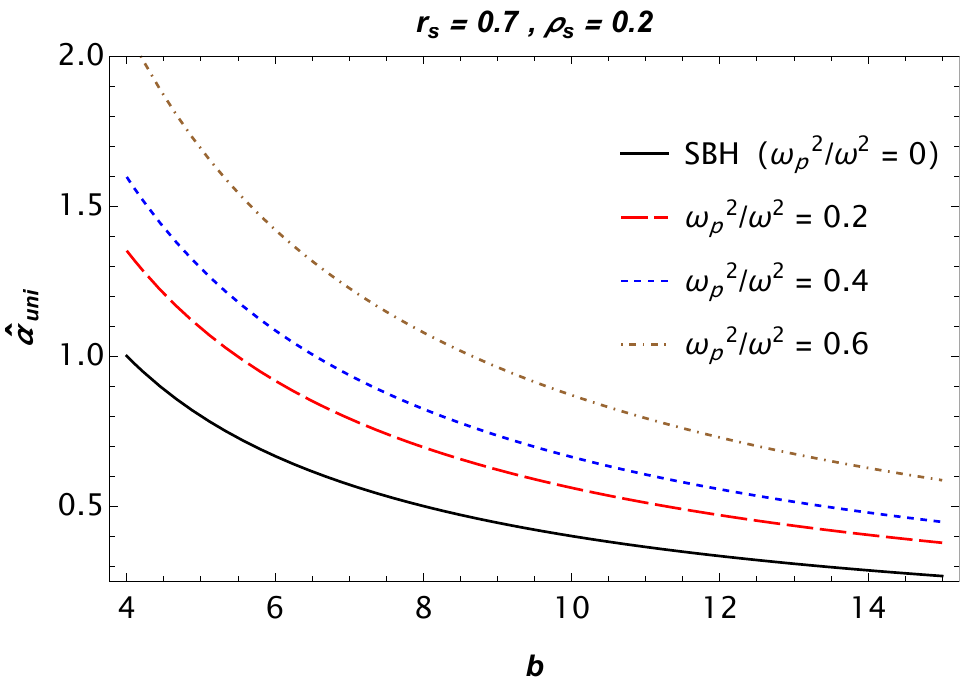}
   \caption{Deflection angle in uniform plasma medium as a function of impact parameter $b$ for different values of parameters.}
    \label{fighomoangle}
\end{figure*}
Using Eq.~\eqref{eqhomoangle}, we compute the results shown in Fig.~\ref{fighomoangle}. The plots indicate that the halo parameters $r_s$ and $\rho_s$, as well as the uniform plasma parameter $\omega_p^2/\omega^2$, have the same qualitative effect: increasing any of these parameters leads to a increase in the deflection angle compared to the SBH case.

\subsection{Singular Isothermal Sphere}

The Singular Isothermal Sphere (SIS) serves as a highly effective framework for characterizing gravitational lensing phenomena involving photons. Originally developed to investigate the properties of lenses and galaxy clusters, the SIS describes a spherical gaseous configuration whose density diverges at the center. Its radial density profile is expressed as \cite{Bisnovatyi-Kogan:2010flt,Babar:2021exh}:

\begin{equation}
\rho(r) = \frac{\sigma_{\nu}^{2}}{2\pi r^{2}},
\label{eq:sis_density}
\end{equation}
where \(\sigma_{\nu}^{2}\) denotes the one-dimensional velocity dispersion. The corresponding particle number density of the plasma is given by the analytic relation \cite{Bisnovatyi-Kogan:2010flt,Babar:2021exh}:

\begin{equation}
N(r) = \frac{\rho(r)}{\kappa m_{p}},
\label{eq:sis_plasma_density}
\end{equation}
with \(m_{p}\) representing the proton mass and \(\kappa\) a dimensionless constant typically associated with the dark matter component of the Universe. Substituting Eq.~\eqref{eq:sis_density} into the definition of the plasma frequency yields:

\begin{equation}
\omega_{e}^{2} = K_{e} N(r) = \frac{K_{e} \sigma_{\nu}^{2}}{2\pi \kappa m_{p} r^{2}}.
\label{eq:sis_plasma_freq}
\end{equation}
Incorporating the aforementioned characteristics of the SIS model, the deflection angle \(\hat{\alpha}_{\text{SIS}}\) takes the form:
\begin{align}\label{sumdefangle}
    \hat{\alpha}_{\text{SIS}} &= \hat{\alpha}_{\text{SIS}1} + \hat{\alpha}_{\text{SIS}2} + \hat{\alpha}_{\text{SIS}3}\ ,\nonumber \\
    \hat{\alpha}_{\text{SIS}1} &= \frac{1}{2} \int_{-\infty}^{\infty} \frac{b}{r} \frac{dh_{33}}{dr} \, dz\ ,\nonumber \\
    \hat{\alpha}_{\text{SIS}2} &= \frac{1}{2} \int_{-\infty}^{\infty} \frac{b}{r} \frac{1}{1 - \omega_c^2 / \omega^2} \frac{d h_{00}}{dr} \, dz\ ,\nonumber \\
    \hat{\alpha}_{\text{SIS}3} &= \frac{1}{2} \int_{-\infty}^{\infty} \frac{b}{r} \left( -\frac{K_e}{\omega^2 - \omega_c^2} \frac{dN}{dr} \right) dz.
\end{align}
where we have introduced an auxiliary plasma constant \(\omega_{c}^{2}\), defined analytically as \cite{Babar:2021exh}:

\begin{equation}
\omega_{c}^{2} = \frac{K_{e} \sigma_{\nu}^{2}}{8\pi \kappa m_{p}}.
\label{eq:omega_c}
\end{equation}

\begin{figure*}
    \centering
    \includegraphics[width=0.45\linewidth]{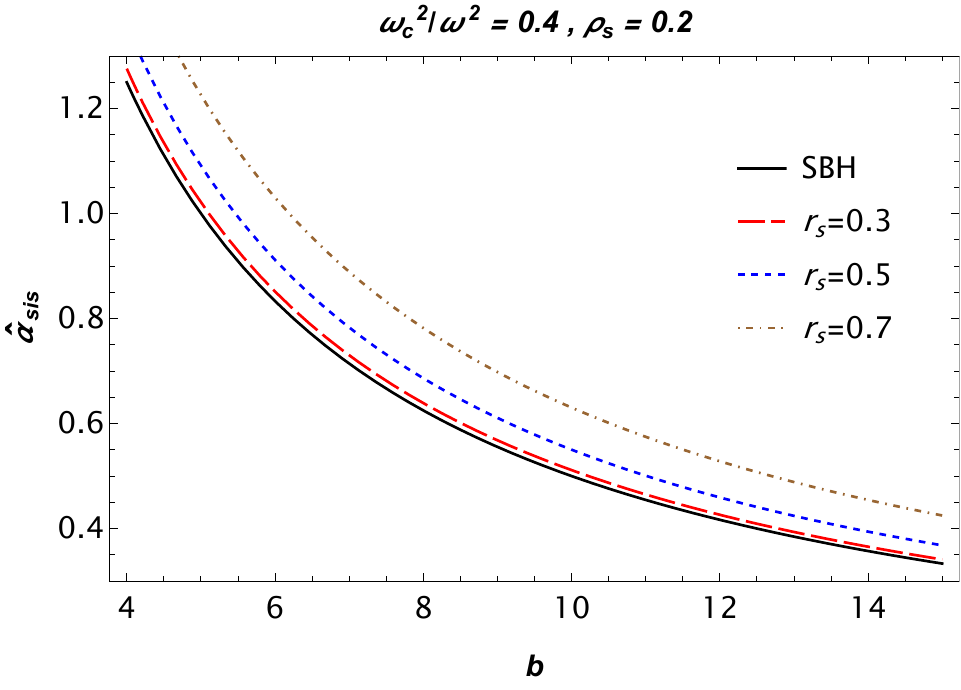}
    \includegraphics[width=0.45\linewidth]{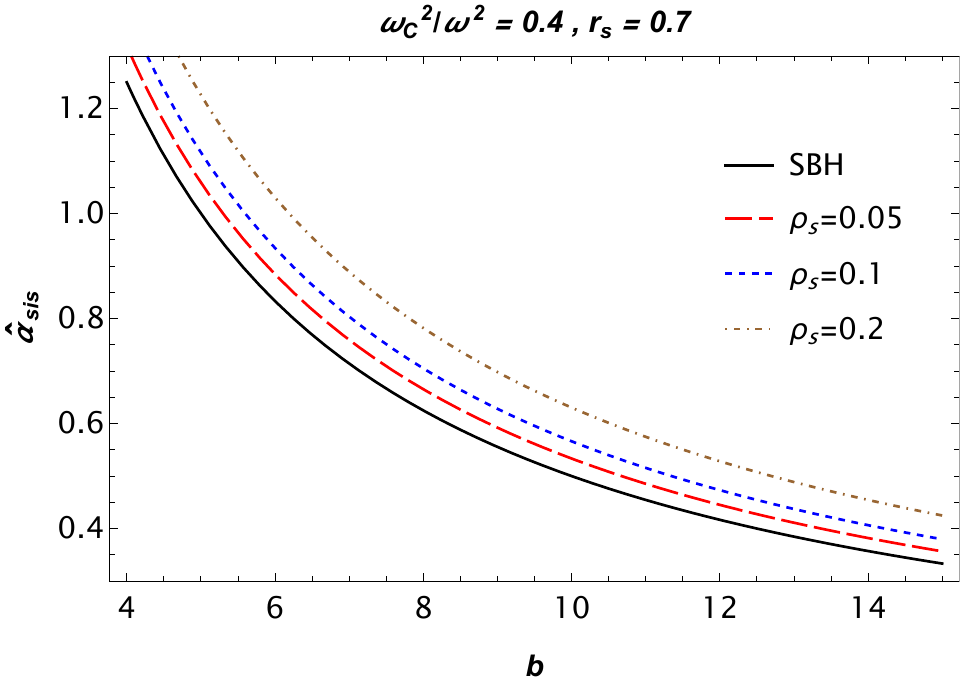}
    \includegraphics[width=0.45\linewidth]{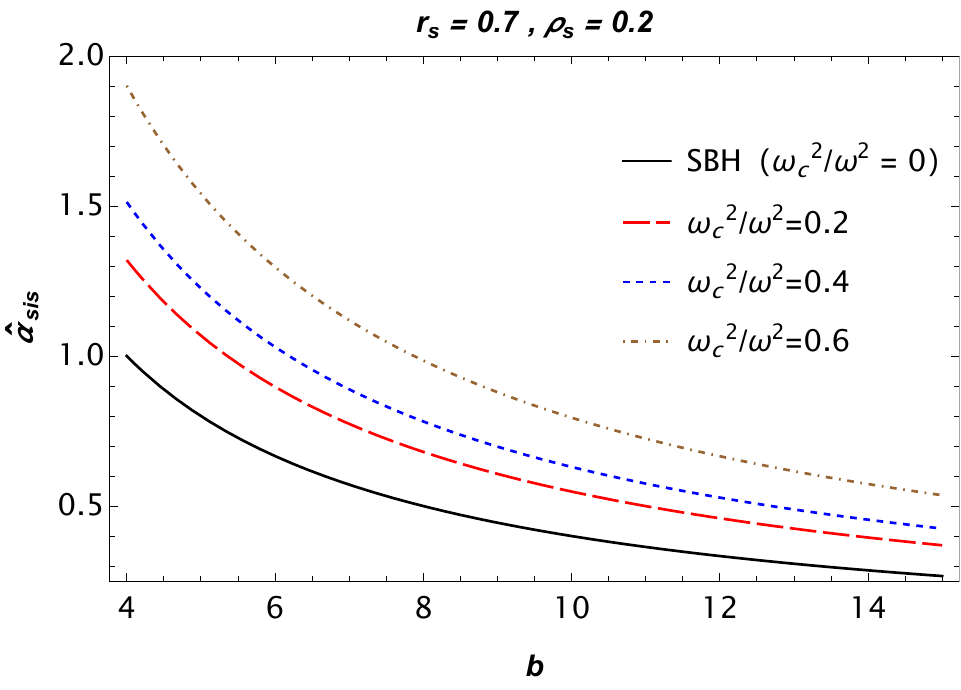}
    \caption{The deflection angle $\alpha_{SIS}$ as function of impact parameter $b$ for different values of parameters.}
    \label{figsisangle}
\end{figure*}

To elucidate the impact of the SIS on photon trajectories, we present the behavior of the deflection angle \(\hat{\alpha}_{SIS}\) as a function of the impact parameter \(b\) in Fig.~\ref{figsisangle}. As observed in Fig.~\ref{fighomoangle}, an increase in the parameters $r_s$ and $\rho_s$ (dark matter halo parameters), as well as in $\omega_c^2/\omega^2$ (non-uniform plasma parameter), leads to an increase in the photon deflection angle in a non-uniform plasma.

\begin{figure}
    \centering
    \includegraphics[width=1\linewidth]{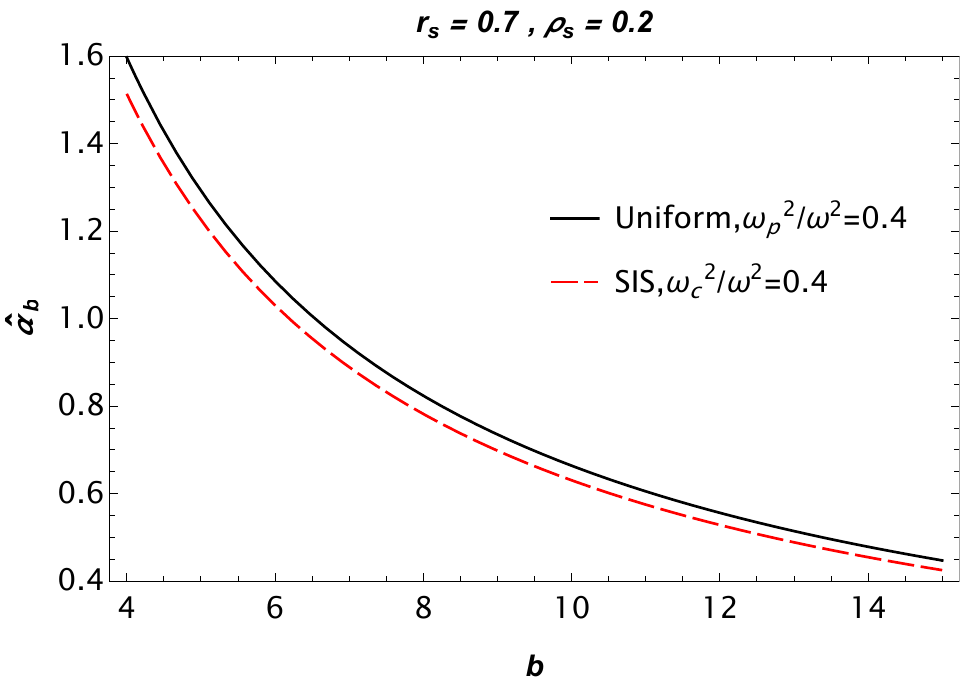}
    \caption{Comparison of the deflection angle for different plasma models under the same set of the parameters.}
    \label{figplasmamodels}
\end{figure}

To illustrate the influence of different plasma distributions, we present in Fig.~\ref{figplasmamodels} a comparative analysis of the deflection angle for plasma models, adopting the same set of physical parameters for consistency. The figure shows that the deflection angle in the SIS model is slightly smaller than the deflection angle of a photon propagating in a uniform plasma.

\section{Magnification of the gravitationally lensed image}\label{sec.6}

Now we investigate the brightness of the image by the deflection angle around the Schwarzschild BH surrounded by a dark matter halo in the presence of the uniform and non-uniform plasma. The combination of light angles can be written as follows by using the lens equation~\cite{10.1111/j.1365-2966.2010.16290.x,PhysRevD.78.103005}
\begin{align}
    \theta D_s=\beta D_s+\hat{\alpha}{D_{ds}}\,,
\end{align}
where $D_\mathrm{s}$ is the distance between the source and the observer, $D_\mathrm{d}$ is the distance between the lens and the observer, and $D_\mathrm{ds}$ is the distance from the source to the lens, while $\theta$ and $\beta$ refer to the angular position of the image and source, respectively. One can find the $\beta$ from the above equation as~\cite{Babar:2021exh}:
\begin{align}\label{newlenseq}
\beta=\theta -\frac{D_\mathrm{ds}}{D_\mathrm{s}}\frac{\xi(\theta)}{D_\mathrm{d}}\frac{1}{\theta}\ .
\end{align}
with
\begin{align}
\xi(\theta)=|\hat{\alpha}_b|b \quad \textrm{and}\quad b=D_\mathrm{d}\theta\,.
\end{align}
It is worth noting that we consider Einstein's ring with radius $R_0=D_d\theta_0$ when the shape of the image looks like a ring. After that, we can find the expression for $\theta_E$ in the following form~\cite{Schneider1992}
\begin{align}
\theta_0=\sqrt{2R_s\frac{D_{ds}}{D_dD_s}}\,.
\end{align}
The magnification of the brightness can be found by using the following equation
\begin{equation}
\mu_{\Sigma} = \frac{I_{\text{tot}}}{I_{*}} = \sum_{k=1}^{j} \left| \left( \frac{\theta_k}{\beta} \right) \left( \frac{d\theta_k}{d\beta} \right) \right|,
\end{equation}
with $I_*$ and $I_{tot}$ representing the unlensed brightness of the source and total brightness, respectively. One can write the magnification of the source as:
\begin{align}
{\mu_{\pm}^{\mathrm{pl}}}=\frac{1}{4}\left ({\frac{x}{\sqrt{x^2+4}}+\frac{\sqrt{x^2+4}}{x} \pm 2}\right )
\end{align}
with $x=\beta/\theta_0$ is the dimensionless quantity~\cite{Babar:2021exh}. The subscript $\pm$ refers to the location of the image with respect to the source and lens, while the superscript $\textrm{pl}$ represents the presence of the plasma. Finally, using the above equations, we can find the total magnification in the following form
\begin{align}\label{mag}
\mu_\mathrm{tot}^{\mathrm{pl}}=\mu_+^{\mathrm{pl}}+\mu^{\mathrm{pl}}_-=\frac{x^2+2}{x\sqrt{x^2+4}}\,.
\end{align}
In the subsequent subsections, we explore the magnification for the uniform and non-uniform plasma cases.

\subsection{Uniform plasma}

\begin{figure*}
    \centering
    \includegraphics[width=0.45\linewidth]{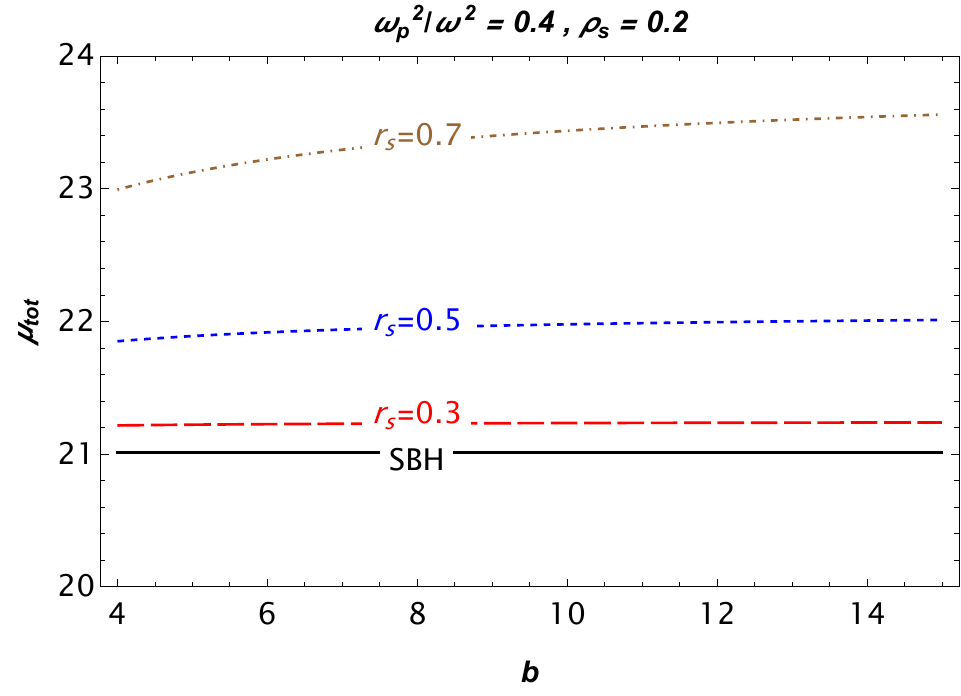}
    \includegraphics[width=0.45\linewidth]{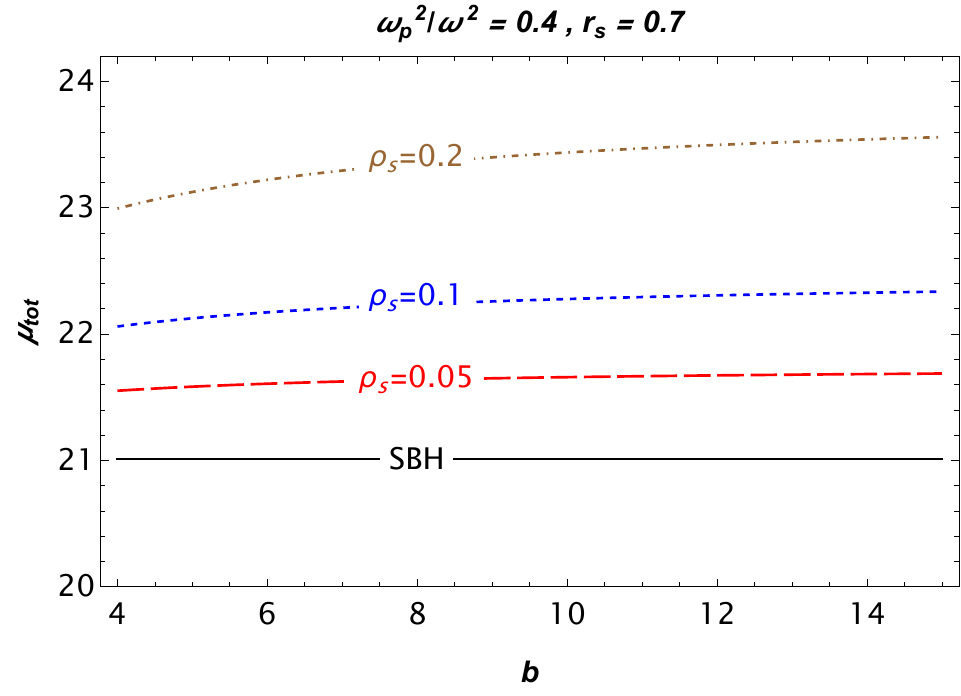}
    \includegraphics[width=0.45\linewidth]{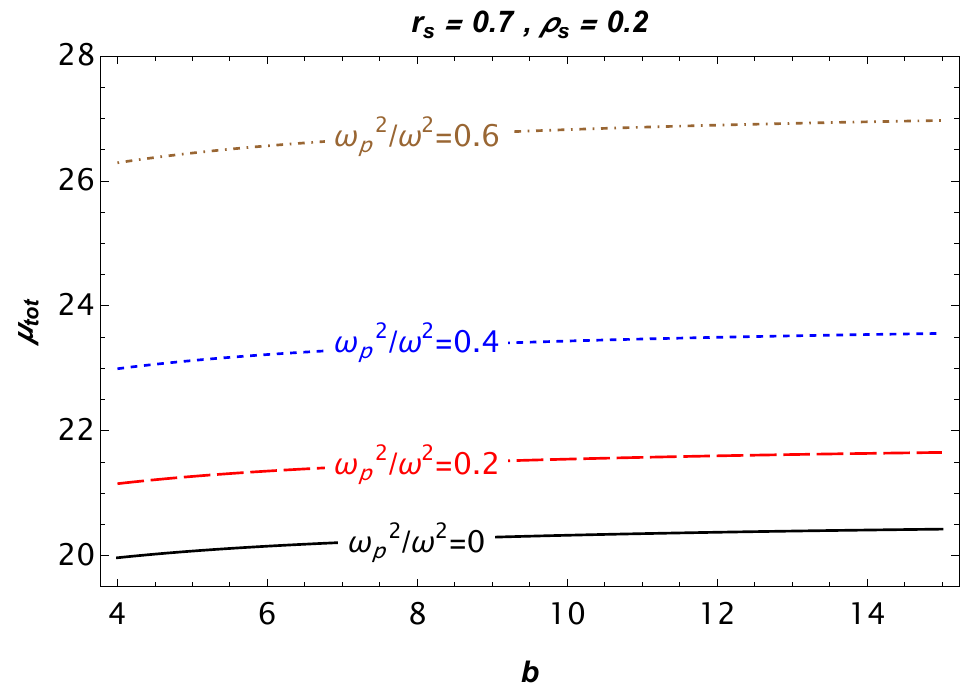}
    \caption{ The plot shows the total magnification as a function of the impact parameter for the different values of the spacetime parameters and uniform plasma frequency.}
    \label{fig:maguni}
\end{figure*}

Now we can rewrite Eq.~(\ref{mag}) for the uniform plasma as follows~\cite{Tsp:2015a}
\begin{equation}
\left(\mu^{\mathrm{pl}}_{\mathrm{tot}}\right)_{\mathrm{uni}}
=
\left(\mu^{\mathrm{pl}}_{+}\right)_{\mathrm{uni}}
+
\left(\mu^{\mathrm{pl}}_{-}\right)_{\mathrm{uni}}
=
\frac{x_{\mathrm{uni}}^{2}+2}{x_{\mathrm{uni}}\sqrt{x_{\mathrm{uni}}^{2}+4}}\,,
\end{equation}
where
\begin{equation}
(\mu^{\mathrm{pl}}_{\pm})_{\mathrm{uni}}=\frac{1}{4}\left(\frac{x_{\mathrm{uni}}}{\sqrt{x_{\mathrm{uni}}^{2}+4}} + \frac{\sqrt{x_{\mathrm{uni}}^{2}+4}}{x_{\mathrm{uni}}} \pm 2 \right)\,,
\end{equation}
with 
\begin{equation}
x_{\mathrm{uni}}=\frac{\beta}{(\theta^{\mathrm{pl}}_0)_{\mathrm{uni}}}\,.
\end{equation}
In Fig.~\ref{fig:maguni}, we plot the total magnification as a function of the impact parameter in the presence of the uniform plasma for the different values of the spacetime parameters and plasma frequency. It can be seen from this figure that the values of the total magnification increase with the rise of the spacetime parameters and the uniform plasma frequency, compared to the SBH case. There is also a slight increase under the influence of the impact parameter.  

\subsection{Singular isothermal sphere}

\begin{figure*}
    \centering
    \includegraphics[width=0.45\linewidth]{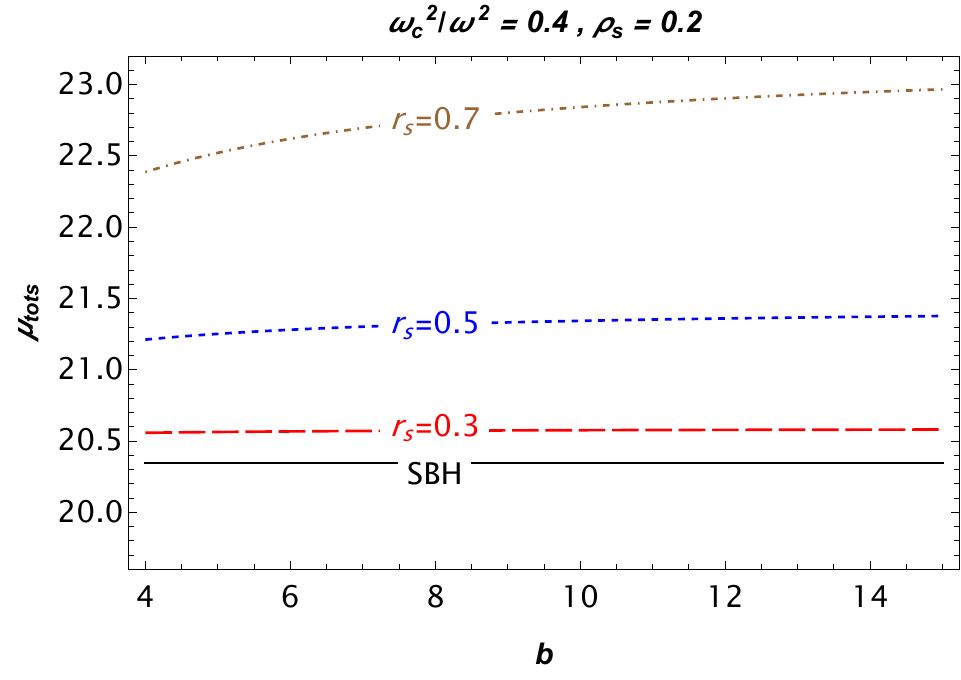}
    \includegraphics[width=0.45\linewidth]{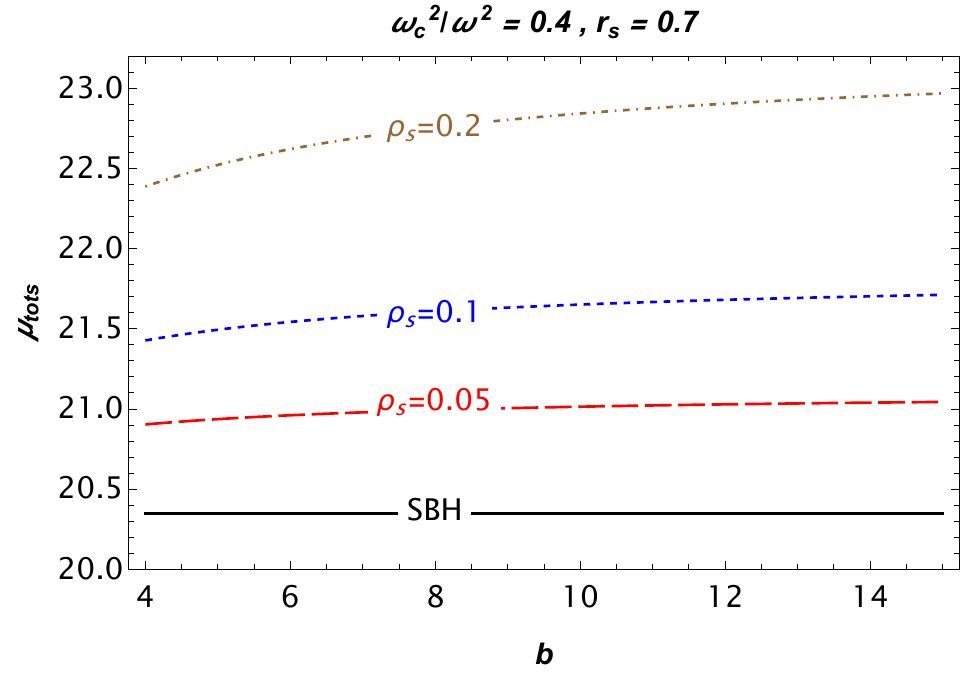}
    \includegraphics[width=0.45\linewidth]{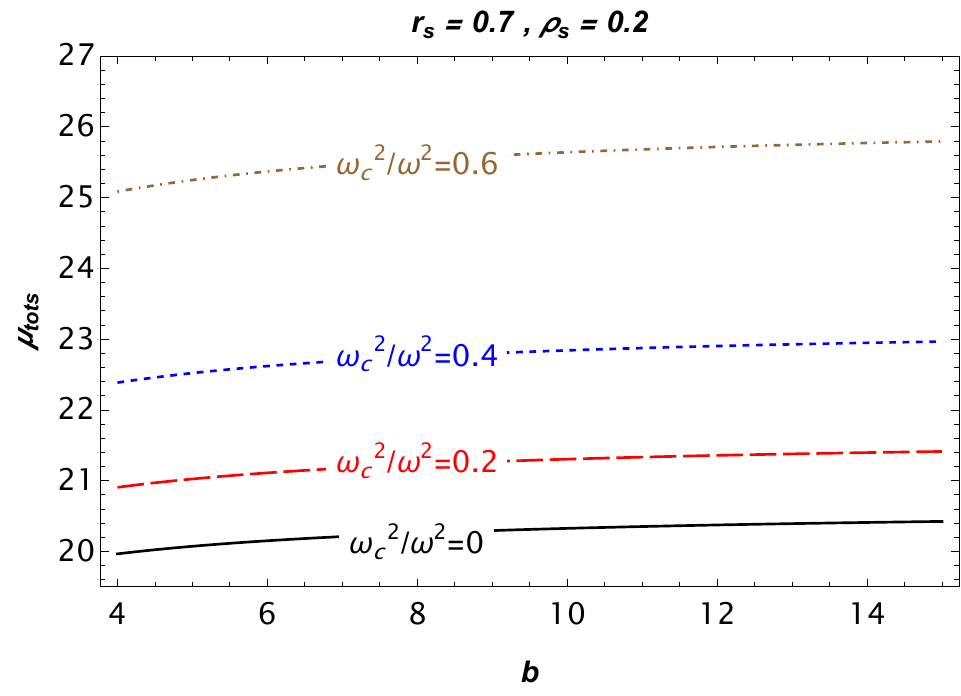}
    \caption{ The plot illustrates the dependence of the total magnification on the impact parameter for the different values of the spacetime parameters and non-uniform plasma frequency. }
    \label{fig:magnonuni}
\end{figure*}

In this subsection, we study the total magnification for the non-uniform plasma case. We carry out the same calculations as in the above part. The total magnification for the non-uniform plasma can be written in the following form~\cite{10.1111/j.1365-2966.2010.16290.x}
\begin{equation}
\left(\mu^{\mathrm{pl}}_{\mathrm{tot}}\right)_{\mathrm{SIS}}
=
\left(\mu^{\mathrm{pl}}_{+}\right)_{\mathrm{SIS}}
+
\left(\mu^{\mathrm{pl}}_{-}\right)_{\mathrm{SIS}}
=
\frac{x_{\mathrm{SIS}}^{2}+2}{x_{\mathrm{SIS}}\sqrt{x_{\mathrm{SIS}}^{2}+4}}\,,
\end{equation}
where
\begin{equation}
(\mu^{\mathrm{pl}}_{\pm})_{\mathrm{SIS}}=\frac{1}{4}\left(\frac{x_{\mathrm{SIS}}}{\sqrt{x_{\mathrm{SIS}}^{2}+4}} + \frac{\sqrt{x_{\mathrm{SIS}}^{2}+4}}{x_{\mathrm{SIS}}} \pm 2 \right)\,,
\end{equation}
and
\begin{equation}
x_{\mathrm{SIS}}=\frac{\beta}{(\theta^{\mathrm{pl}}_0)_{\mathrm{SIS}}}\,.
\end{equation}
The Fig.~\ref{fig:magnonuni} demonstrates the dependence of the total magnification on the impact parameter for the different values of the spacetime parameters and the frequency of the non-uniform plasma. Similar to the uniform case, the values of the total magnification increase with the increase of the spacetime parameters and non-uniform plasma frequency. It is worth noting that there is almost no effect of the impact parameter on the total magnification in the  SBH case, while there is a slight rise under the influence of the impact parameter. 
\begin{figure*}
    \centering
    \includegraphics[width=0.32\linewidth]{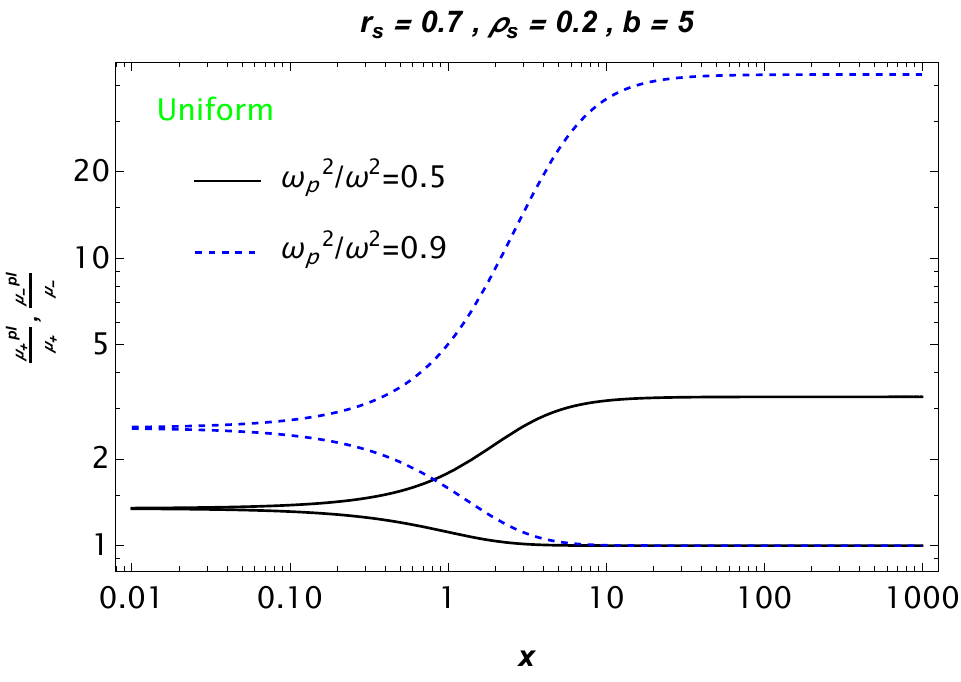}
    \includegraphics[width=0.32\linewidth]{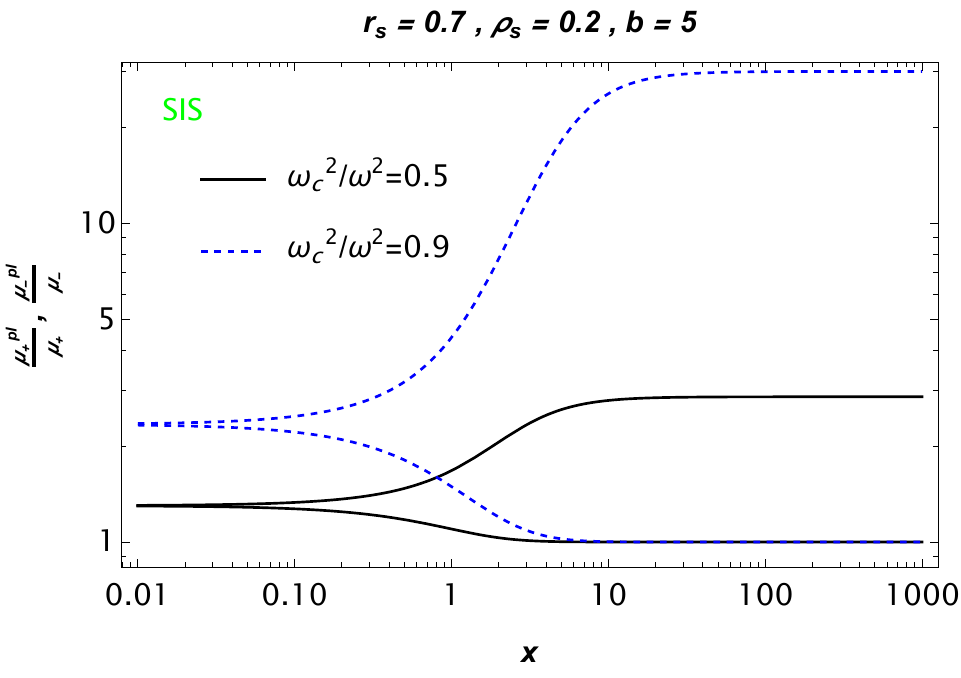}
    \includegraphics[width=0.32\linewidth]{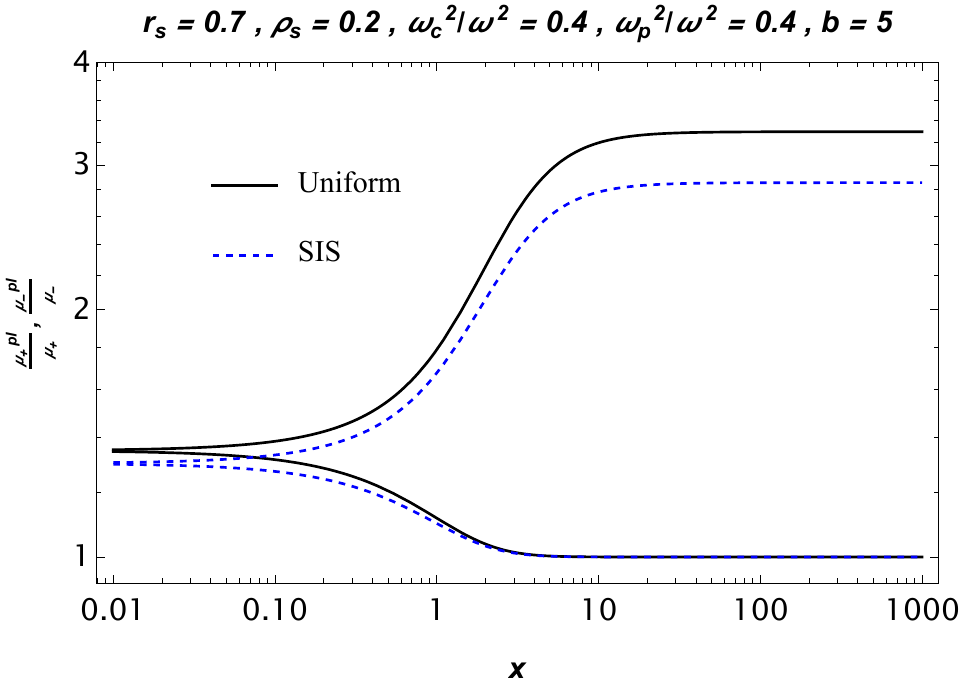}
    \caption{The plot demonstrates the magnification ratio of uniform and non-uniform plasma to vacuum for different values of the plasma frequency.}
    \label{figmagdif}
\end{figure*}
To be more informative, Fig.~\ref{figmagdif} presents the magnification ratio for both uniform and non-uniform plasma relative to the vacuum case. As shown in this figure, all curves converge at a single point as $x_0$ approaches zero. Additionally, we include a comparison between the results obtained for uniform and non-uniform plasma cases.

\section{Black Hole Shadow in a Plasma Environment}\label{sec.7}

We now investigate the shadow radius of a Schwarzschild black hole embedded within a dark matter halo and surrounded by a plasma medium. The angular radius of the black hole shadow, denoted \(\alpha_{sh}\), can be derived using a geometric approach, leading to the expression \cite{Perlick:2015vta,Atamurotov:2021cgh}:
\begin{equation}
    \sin^{2} \alpha_{sh} = \frac{h^{2}(r_{ph})}{h^{2}(r_{0})} =
    \frac{
        r_{ph}^{2} \left[ \dfrac{1}{f(r_{ph})} - \dfrac{\omega_{p}^{2}(r_{ph})}{\omega_{0}^{2}} \right]
    }{
        r_{0}^{2} \left[ \dfrac{1}{f(r_{0})} - \dfrac{\omega_{p}^{2}(r_{0})}{\omega_{0}^{2}} \right]
    },
    \label{eq:shadow_angle}
\end{equation}
where \(r_{0}\) and \(r_{ph}\) denote the positions of the observer and the photon sphere, respectively.

If the observer is situated at a sufficiently large distance from the black hole, the shadow radius can be approximated as \cite{Perlick:2015vta}

\begin{equation}
    R_{sh} \approx r_{0} \sin \alpha_{sh} = r_{ph}^{2} \sqrt{ \frac{1}{f(r_{ph})} - \frac{\omega_{p}^{2}(r_{ph})}{\omega_{0}^{2}} }.
    \label{eq:shadow_radius_approx}
\end{equation}
This approximation follows from the asymptotic behavior \(h(r) \to r\) at spatial infinity, as implied by Eq.~\eqref{funchr} for the considered plasma models.

In the vacuum limit where \(\omega_{p}(r) \equiv 0\), and with the photon sphere radius \(r_{ph} = 3M\), Eq.~\eqref{eq:shadow_radius_approx} reduces to the well-known Schwarzschild shadow radius:

\begin{align}
    R_{\text{sh}} = 3\sqrt{3}M.
\end{align}

\begin{figure*}
    \centering
    \includegraphics[width=0.45\linewidth]{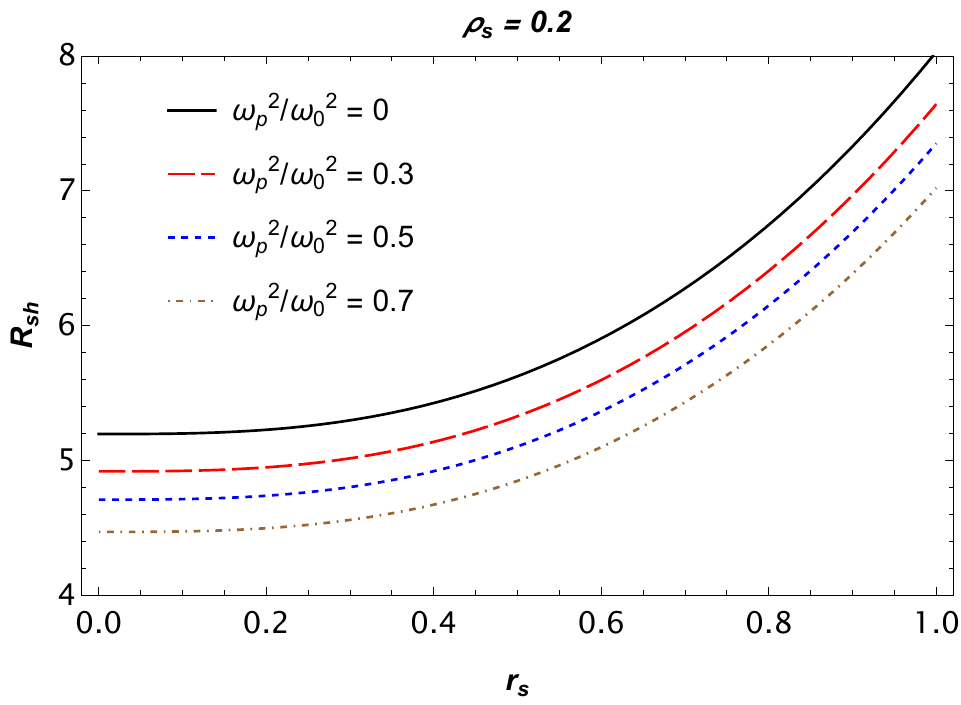}
    \includegraphics[width=0.45\linewidth]{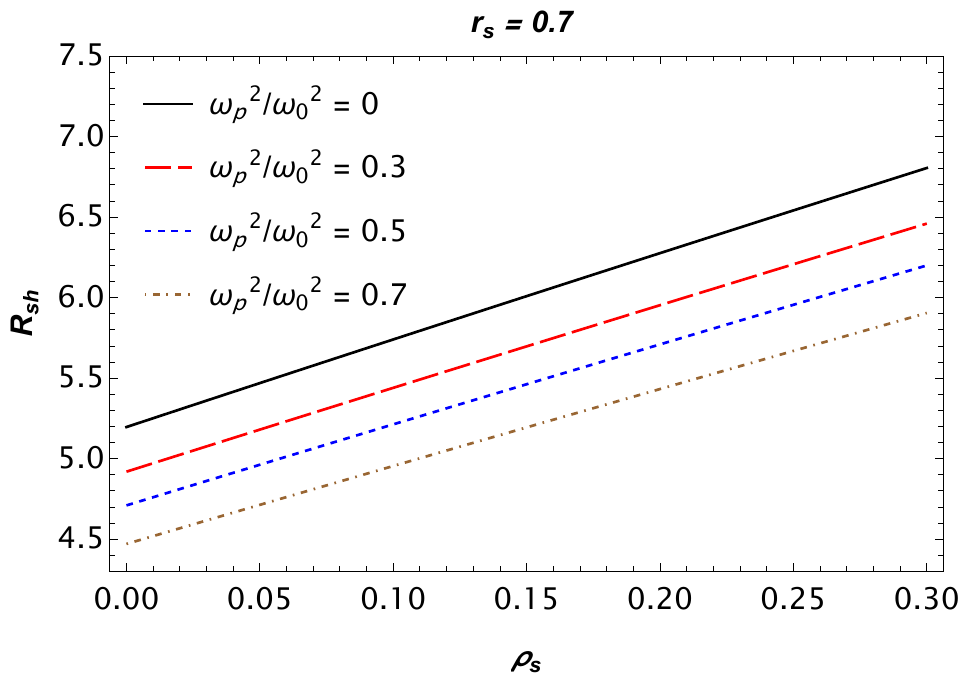}
    \caption{Shadow radius in uniform plasma as a function of the parameters.}
    \label{fighomoshadow}
\end{figure*}

The shadow radius in homogeneous plasma is presented in Fig.~\ref{fighomoshadow}. We observe that the effect of the halo parameters leads to an increase in the shadow radius, whereas the influence of the plasma acts in the opposite direction, reducing its size.

\begin{figure*}
    \centering
    \includegraphics[width=0.45\linewidth]{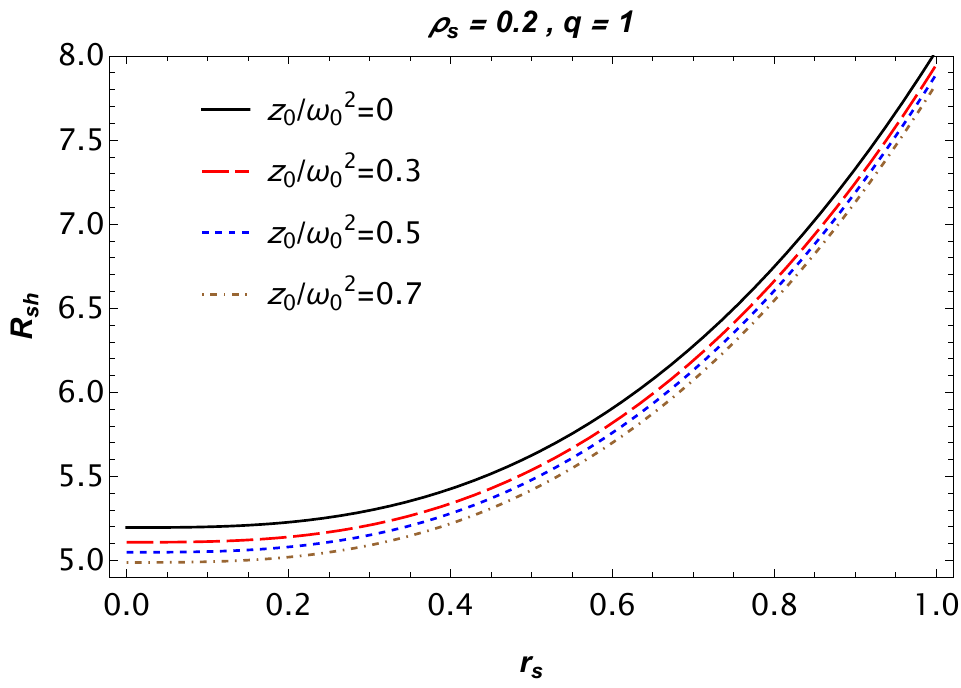}
    \includegraphics[width=0.45\linewidth]{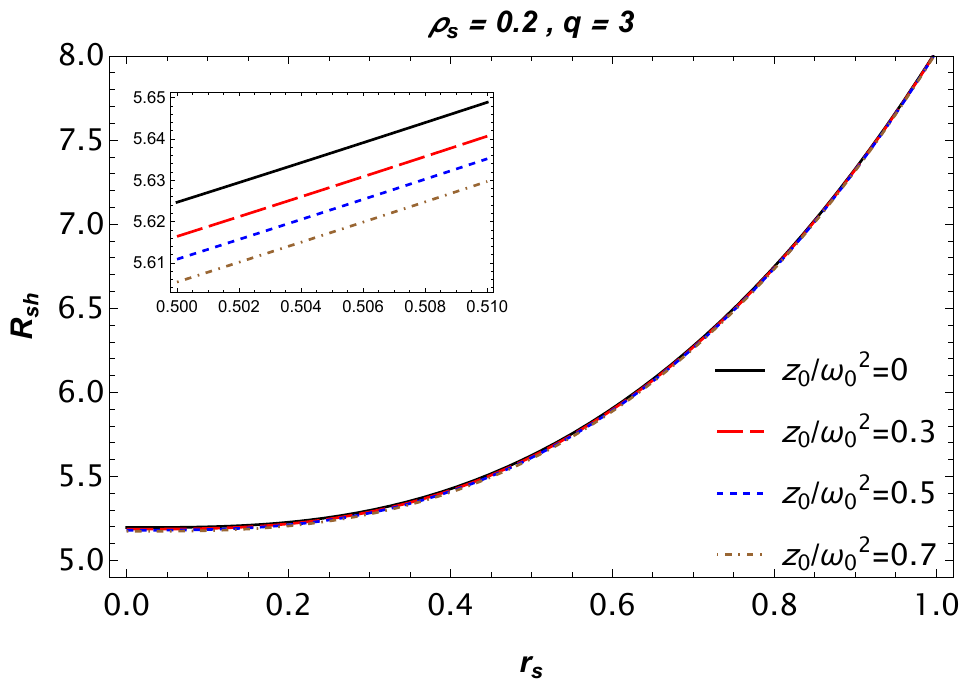}
    \includegraphics[width=0.45\linewidth]{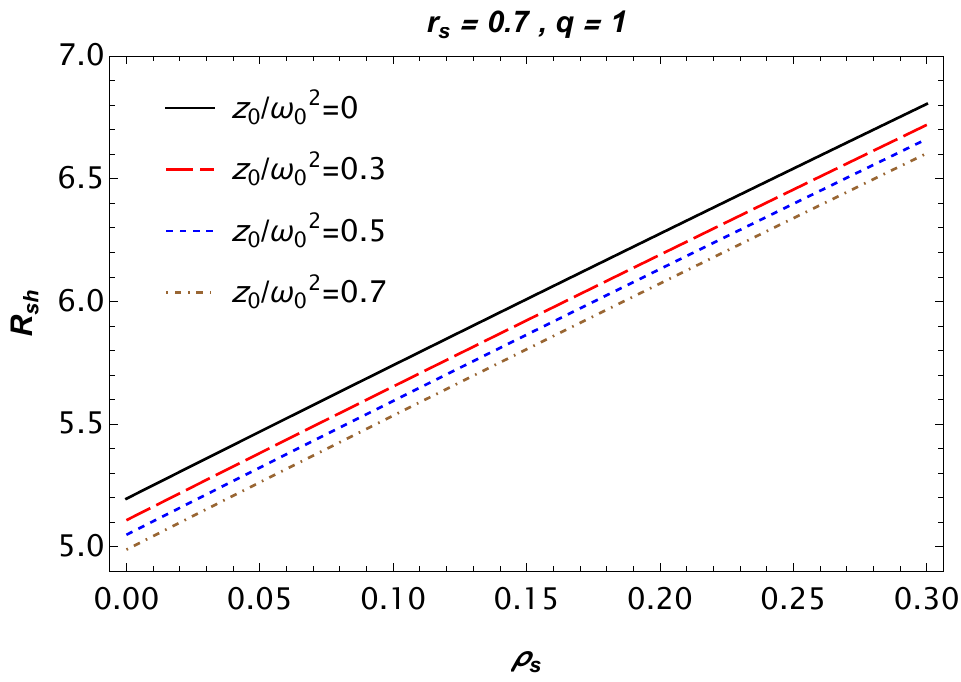}
    \includegraphics[width=0.45\linewidth]{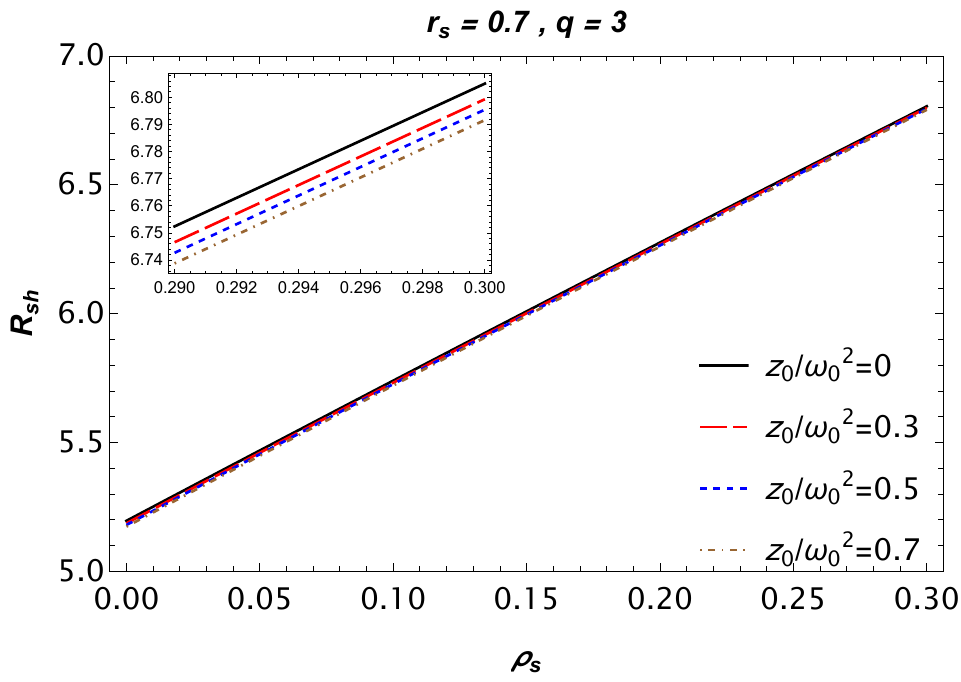}
    \caption{Shadow radius in non-uniform plasma as a function of the parameters.}
    \label{fignonhomoshadow}
\end{figure*}

The shadow radius in a nonhomogeneous plasma is shown in Fig.~\ref{fignonhomoshadow}. The impact of the halo parameters is similar to that in the homogeneous plasma model. Additionally, the effect of the plasma becomes negligible when its parameter is $q = 3$.

\section{Parameter Estimation via MCMC Analysis}\label{sec.8}

We employed a Markov Chain Monte Carlo (MCMC) sampling technique, implemented using the \texttt{emcee}\cite{Foreman-Mackey:2012any} Python package, to estimate the mass and plasma parameters of the Schwarzschild black hole within a dark matter halo. This method samples from the posterior probability distribution of the model parameters, conditioned on observational data.

The likelihood function was constructed using the measured angular shadow diameters of \(\text{M87}^*\) and \(\text{Sgr A}^*\) provided by the Event Horizon Telescope (EHT) collaboration, as summarized in Table~\ref{table1}. The theoretical shadow diameter in our model is linked to the radius of the photon sphere, which is obtained by numerically solving the corresponding null geodesic equation.

\begin{table*}[]
    \centering
    \caption{Observational parameters for \(\text{M87}^*\) and \(\text{Sgr A}^*\) \cite{EventHorizonTelescope:2024dhe,DeLaurentis:2022nrv,GRAVITY:2020gka}.}
    \begin{tabular}{c|c|c}
        \hline
        Parameter & \(\text{M87}^*\) & \(\text{Sgr A}^*\) \\
        \hline
        Angular Diameter (\(\theta\)) & \(43.3 \pm 2.3\ \mu\text{as}\) & \(51.8 \pm 2.3\ \mu\text{as}\) \\
        Distance (\(D\)) & \(16.5\ \text{Mpc}\) & \(8.275\ \text{kpc}\) \\
        Mass (\(M\)) & \((6.5 \pm 0.7) \times 10^9 M_{\odot}\) & \((4.297 \pm 0.013) \times 10^6 M_{\odot}\) \\
        \hline
    \end{tabular}
    \label{table1}
\end{table*}

The predicted angular size of the shadow, \(\theta_{\rm sh}\), is calculated from the theoretical shadow radius \(R_{\rm sh}\) via:

\begin{widetext}
    \begin{align}
    \theta_{\rm sh, rad} &= \frac{2 R_{\rm sh} G M}{c^2 D_{\rm obs}}, \nonumber \\
    \theta_{\rm sh, \mu as} &= \theta_{\rm sh, rad} \times \frac{180 \times 3600 \times 10^6}{\pi} = 2.06265 \times 10^{11} \times \theta_{\rm sh, rad}, \label{eq:theta_sh}
\end{align}
\end{widetext}
where \(G = 6.67 \times 10^{-11}\ \text{m}^3\ \text{kg}^{-1}\ \text{s}^{-2}\) is Newton's constant, \(c = 3 \times 10^8\ \text{m}\ \text{s}^{-1}\) is the speed of light, and \(D_{\rm obs}\) is the distance to the black hole (taken as \(16.5\ \text{Mpc}\) for \(\text{M87}^*\) and \(8.275\ \text{kpc}\) for \(\text{Sgr A}^*\)).

The log-likelihood function is defined as:
\begin{equation}
    \log \mathcal{L}(\boldsymbol{\Theta}) = -\frac{1}{2} \sum_i \frac{\left( \theta_{{\rm sh}, i}^{\rm pred}(\boldsymbol{\Theta}) - \theta_{{\rm sh}, i}^{\rm obs} \right)^2}{\sigma_i^2}, \label{eq:log_likelihood}
\end{equation}
where \(\theta_{{\rm sh}, i}^{\rm obs}\) and \(\sigma_i\) are the observed shadow size and its uncertainty, respectively, and \(\boldsymbol{\Theta}\) denotes the set of model parameters.

According to Bayes' theorem, the posterior distribution is proportional to the product of the likelihood and the prior:
\begin{equation}
    P(\boldsymbol{\Theta} | \mathcal{D}) \propto \mathcal{L}(\mathcal{D} | \boldsymbol{\Theta}) \cdot \pi(\boldsymbol{\Theta}), \label{eq:posterior}
\end{equation}
where \(\pi(\boldsymbol{\Theta})\) represents the prior probability distribution. We adopted Gaussian priors for the black hole mass, centered on the literature values, and uniform priors for the other parameters within physically plausible ranges:

\noindent For \(\text{M87}^*\):
\begin{equation}
\begin{cases}
6.0 \times 10^9 M_\odot < M < 10 \times 10^9 M_\odot \\
-1.0  < Q/M < 1.0 \\
0  < z_0/(M\omega_0^2) < 1.0 \\
0 < q < 5
\end{cases}
\label{eq:priors_m87}
\end{equation}

\noindent For \(\text{Sgr A}^*\):
\begin{equation}
\begin{cases}
4.0 \times 10^6 M_\odot < M < 4.6 \times 10^6 M_\odot \\
-1.0  < Q/M < 1.0 \\
0  < z_0/(M\omega_0^2) < 1.0 \\
0 < q < 5
\end{cases}
\label{eq:priors_sgr}
\end{equation}

\begin{figure*}
    \centering
   \includegraphics[width=0.45\linewidth]{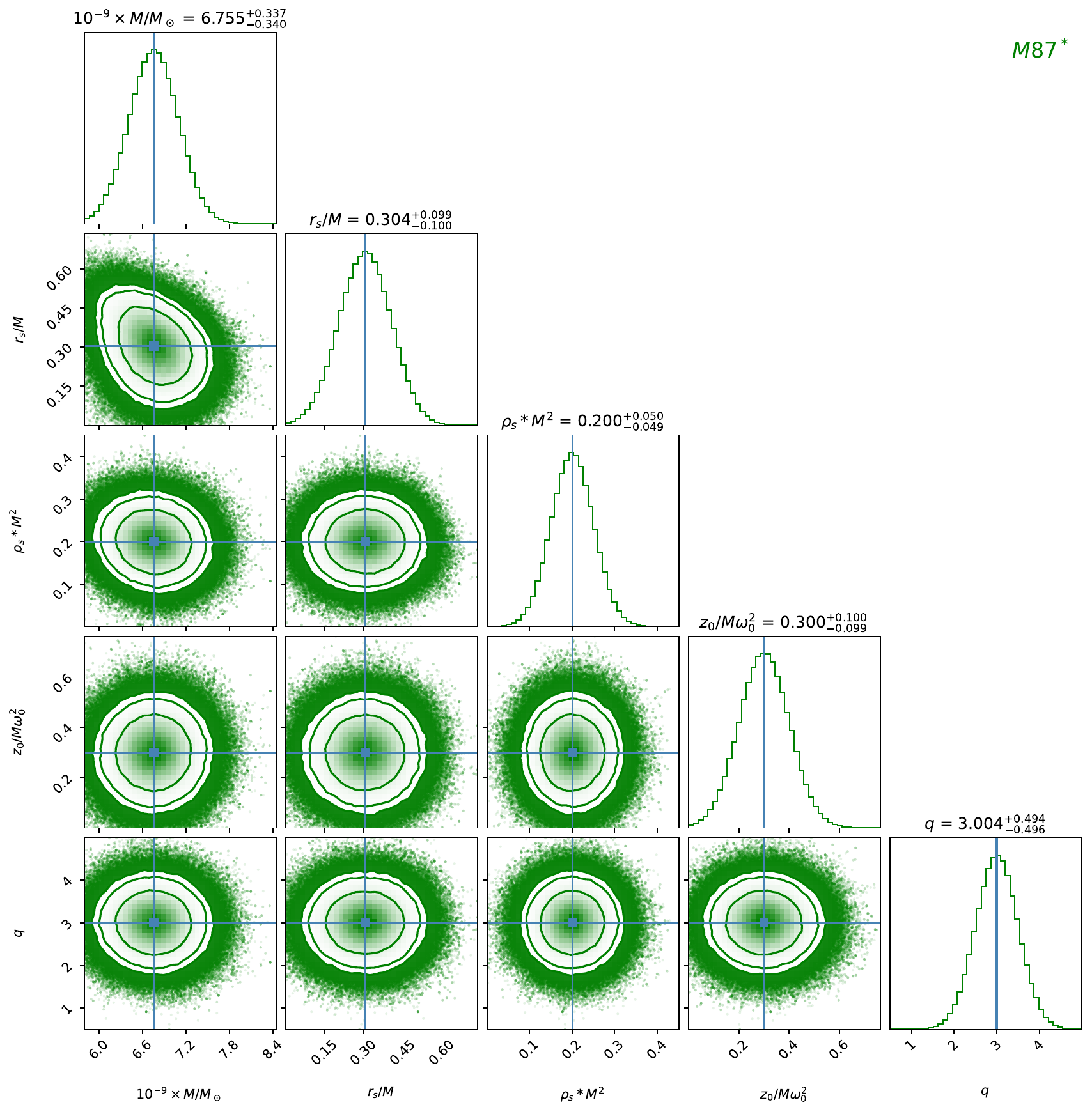}
   \includegraphics[width=0.45\linewidth]{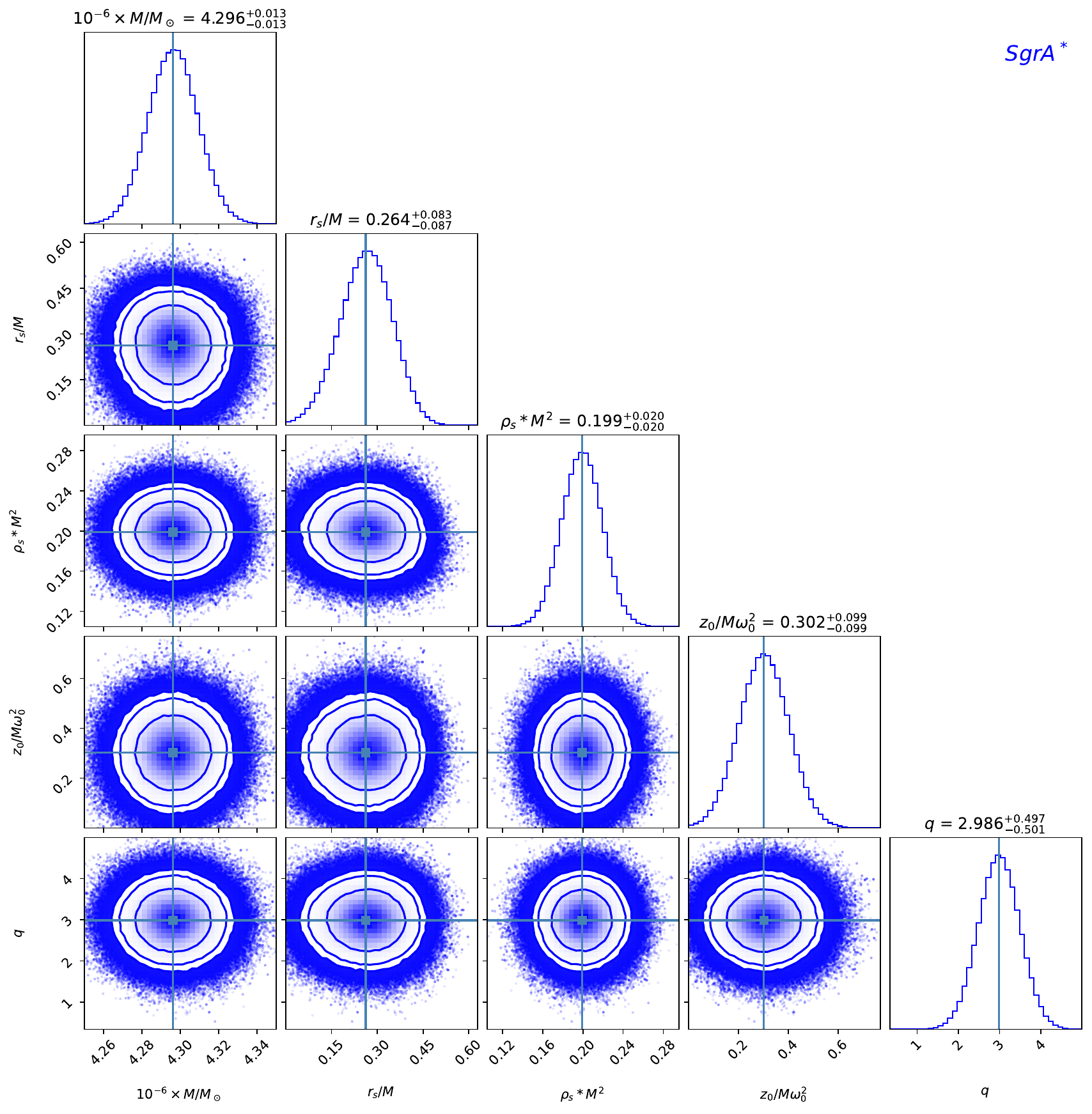}
    \caption{The plot shows the estimated values of BH mass, DM parameters and the plasma frequency by using the EHT data for $\text{M87}^*$ (left panel) and $\text{Sgr A}^*$ (right panel). }
    \label{mcmc_shadow}
\end{figure*}
The MCMC analysis, whose sampling results are visualized in Fig.~\ref{mcmc_shadow}, yields estimates for the mass, dar matter halos' parameters, and plasma coefficients for both \(\text{M87}^*\) and \(\text{Sgr A}^*\). The best-fit parameter values, presented in Table~\ref{table2}, demonstrate consistency with the EHT observational constraints, supporting the physical viability of our plasma-modified gravitational model.

\begin{table*}[]
    \centering
    \caption{Best-fit parameter values for the plasma-corrected black hole model, inferred from the shadow data for \(\text{M87}^*\) and \(\text{Sgr A}^*\).}
    \begin{tabular}{c|c|c}
        \hline
        Parameter & \(\text{M}87^*\) & \(\text{Sgr A}^*\) \\
        \hline
        \(M\) & \(6.755^{+0.337}_{-0.340} \times 10^9 M_{\odot}\) & \(4.296 \pm 0.013 \times 10^6 M_{\odot}\) \\
        \(r_s/M\) & \(0.304^{+0.099}_{-0.100}\) & \(0.264^{+0.083}_{-0.087}\) \\
        \(\rho_s*M^2\) & \(0.200^{+0.050}_{-0.049}\) & \(0.199\pm0.020\) \\
        \(z_0/(M\omega_0^2)\) & \(0.300^{+0.100}_{-0.099}\) & \(0.302 \pm 0.099\) \\
        \(q\) & \(3.004^{+0.494}_{-0.496}\) & \(2.986^{+0.497}_{-0.501}\) \\
        \hline
    \end{tabular}
    \label{table2}
\end{table*}

\section{Conclusions}\label{sec.9}

In this article, we investigate the observational properties of the Schwarzschild BH surrounded by a dark matter halo, including the constraint of the spacetime parameters by using the EHT results. One can summarize our main results as :
\begin{itemize}
    \item Our study commences with a brief review of the spacetime of the Schwarzschild BH surrounded by a dark matter halo. We investigate the structure of the event horizon. It is found that increasing the halo parameters leads to an expansion of the event horizon radius; in other words, larger values of the halo parameters result in a larger horizon. 
    \item Furthermore, using the Lagrangian formalism, we explore the massive particle motion around the Schwarzschild BH surrounded by a dark matter halo. We derive the analytic expression for the effective potential of the massive particle and plot the radial dependence of the effective potential, presenting the maximum and minimum values for different spacetime parameters. It was found that the values of the effective potential decrease with the rise of the spacetime parameters. After that, the analytic expressions for the specific energy and angular momentum of a massive particle in circular motion were derived by analyzing the effective potential, and plotting their behaviour. Additionally, by applying the corresponding conditions, we determine the ISCO radii. It was found that the values of the ISCO radius increase with the increase of the dark matter halo parameters.    
    \item Moreover, we investigate the photon motion around the BH in the presence of the homogeneous and inhomogeneous plasma using the Hamiltonian formalism. We find the radius of the photon sphere for each case, and we plot them as a function of the spacetime parameters. It can be seen from their behaviours that the photon sphere radii increase with the rise of spacetime parameters and the homogeneous plasma frequency. The impact of the spacetime parameters in the inhomogeneous case is the same, but the plasma frequency slightly increases the values of the photon sphere radii.  
    \item In addition, we study the gravitational weak lensing around the Schwarzschild BH surrounded by a dark matter halo in the presence of the uniform and non-uniform plasma. Note that we perform numerical calculations to analyze the impact of the spacetime parameters and the plasma frequency on the deflection angle for each case due to the complex spacetime. We found that the values of the deflection angle for both cases increase under the influence of the spacetime parameters and plasma frequency.  To provide a more informative comparison, we fix other parameters and examine the deflection angles. Additionally, we explore the magnification of the gravitationally lensed image. We plot the total magnification as a function of the impact parameter for the different values of the plasma frequency and spacetime parameters. It was found that the spacetime parameters and plasma frequency increase the values of the total magnification in both cases. Subsequently, we compare the magnification ratio for both cases relative to the vacuum to be more informative.
    \item Finally, we study the BH shadow using the same formalism as the investigation of photon sphere radii for homogeneous and inhomogeneous plasma environments. The results indicate that the values of the shadow radii decrease with the rise of the plasma frequency and vice versa for the spacetime parameters. We then constrained the spacetime parameters using the observational data released by the EHT collaboration.  
\end{itemize}

\section*{ACKNOWLEDGEMENT}

This research was funded by the National Natural Science Foundation of China (NSFC) under Grant No. U2541210.

\bibliographystyle{apsrev4-1} 
\bibliography{Ref}

\end{document}